\makeatletter
\@ifundefined{@parse@version@dash}{%
\def\@parse@version#1{\@parse@version@0#1}
\def\@parse@version@#1/#2/#3#4#5\@nil{%
\@parse@version@dash#1-#2-#3#4\@nil}
\def\@parse@version@dash#1-#2-#3#4#5\@nil{%
  \if\relax#2\relax\else#1\fi#2#3#4 }
}{}
\makeatother

\documentclass[twocolumn,showpacs,showkeys,preprintnumbers,amsmath,amssymb,superscriptaddress]{revtex4-2}
\usepackage{graphicx}
\usepackage{natbib}
\usepackage{amsmath}
\usepackage{epsfig}
\usepackage{amssymb}
\usepackage{xcolor}
\usepackage{cancel}
\usepackage{ulem}
\usepackage{lineno}
\usepackage{hyperref}
\hypersetup{
 colorlinks,
 linkcolor={red},
 citecolor={blue},
 urlcolor={blue}
}
\hypersetup{breaklinks=true}

\begin{document}


\title{Hindered Prompt-Neutron Evaporation in Surrogate Reactions for $^{239}$Pu(n,f)}

\author{D.~Ramos}
\email[]{diego.ramos@ganil.fr}
\affiliation{GANIL, CNRS/IN2P3, CEA/DRF, bd Henri Becquerel, 14076 Caen, France}

\author{M.~Caama\~no}
\email[]{manuel.fresco@usc.es}
\affiliation{IGFAE - Universidade de Santiago de Compostela, E-15706 Santiago de Compostela, Spain}

\author{F.~Farget}
\affiliation{GANIL, CNRS/IN2P3, CEA/DRF, bd Henri Becquerel, 14076 Caen, France}

\author{C.~Rodr\'iguez-Tajes}
\affiliation{GANIL, CNRS/IN2P3, CEA/DRF, bd Henri Becquerel, 14076 Caen, France}

\author{A.~Lemasson}
\affiliation{GANIL, CNRS/IN2P3, CEA/DRF, bd Henri Becquerel, 14076 Caen, France}

\author{M.~Rejmund}
\affiliation{GANIL, CNRS/IN2P3, CEA/DRF, bd Henri Becquerel, 14076 Caen, France}

\author{C.~Schmitt}
\altaffiliation[Present address: ]{IPHC Strasbourg, Universit\'e de Strasbourg CNRS/IN2P3, F-67037 Strasbourg Cedex 2, France}
\affiliation{GANIL, CNRS/IN2P3, CEA/DRF, bd Henri Becquerel, 14076 Caen, France}

\author{E.~Clement}
\affiliation{GANIL, CNRS/IN2P3, CEA/DRF, bd Henri Becquerel, 14076 Caen, France}

\author{O.~Litaize}
\affiliation{CEA, DES, IRESNE, DER, SPRC, LEPh, 13108, Saint Paul Lez Durance, France}

\author{O.~Serot}
\affiliation{CEA, DES, IRESNE, DER, SPRC, LEPh, 13108, Saint Paul Lez Durance, France}

\author{L.~Audouin}
\affiliation{IJC Lab, Universit\'e Paris-Saclay, CNRS/IN2P3, F-91405 Orsay Cedex, France}

\author{J.~Benlliure}
\altaffiliation[Present address: ]{IFIC, Centro Mixto Universidad de Valencia-CSIC, Valencia, Spain}
\affiliation{IGFAE - Universidade de Santiago de Compostela, E-15706 Santiago de Compostela, Spain}

\author{E.~Casarejos}
\affiliation{CINTECX, Universidade de Vigo, DSN, Dpt. Mech. Engineering, E-36310 Vigo, Spain}

\author{D.~Cortina}
\altaffiliation[Present address: ]{IFIC, Centro Mixto Universidad de Valencia-CSIC, Valencia, Spain}
\affiliation{IGFAE - Universidade de Santiago de Compostela, E-15706 Santiago de Compostela, Spain}

\author{D.~Dor\'e}
\affiliation{CEA Saclay, DRF/IRFU/SPhN, 91191 Gif-sur-Yvette Cedex, France}

\author{B.~Fern\'andez-Dom\'inguez}
\affiliation{IGFAE - Universidade de Santiago de Compostela, E-15706 Santiago de Compostela, Spain}

\author{G.~de~France}
\affiliation{GANIL, CNRS/IN2P3, CEA/DRF, bd Henri Becquerel, 14076 Caen, France}

\author{A.~Heinz}
\affiliation{Chalmers University of Technology, SE-41296 G\"oteborg, Sweden}

\author{B.~Jacquot}
\affiliation{GANIL, CNRS/IN2P3, CEA/DRF, bd Henri Becquerel, 14076 Caen, France}

\author{C.~Paradela}
\altaffiliation[Present address: ]{EC-JRC, Institute for Reference Materials and Measurements, Retieseweg 1111, B-2440 Geel, Belgium}
\affiliation{IGFAE - Universidade de Santiago de Compostela, E-15706 Santiago de Compostela, Spain}

\author{T.~Roger}
\affiliation{GANIL, CNRS/IN2P3, CEA/DRF, bd Henri Becquerel, 14076 Caen, France}

\date{\today}

\begin{abstract}

Isotopic fission-fragment distributions of $^{240}$Pu have been measured, for the first time, as a function of the initial excitation energy, and the prompt neutron multiplicity has been derived from these data. The $^{240}$Pu fissioning system was produced through the two-proton transfer reaction between $^{238}$U and $^{12}$C, a surrogate reaction for the neutron-capture-induced fission $^{239}$Pu(n,f). The reaction was measured in inverse kinematics, allowing the fission fragments to be fully identified with the VAMOS Spectrometer. When compared to neutron-capture-induced reactions, the observed prompt neutron multiplicity shows a clear reduction in the surrogate two-proton transfer, revealing an unexpected influence of the entrance channel in the fission output. At the same time, fission-fragment yield distributions obtained in neutron-capture-induced reactions show a relative fission-fragment production in the symmetry region similar to that measured in this work. The discrepancy in neutron multiplicity is attributed to the additional angular momentum induced in the multi-nucleon transfer reactions, which excites the fissioning system to higher-spin states, increasing the probability of gamma emission that competes with neutron evaporation, in particular from the fission barrier to the scission point. This observation underlines the limitations in the utilisation of properties derived from surrogate reactions in nuclear technology and other applications of nuclear fission.
\end{abstract}


\pacs {}

\maketitle


Fission has been the subject of extensive research, primarily through neutron-capture-induced reactions involving long-lived
actinide isotopes. The practical application of this research is directly linked to the production of energy~\cite{NEA}.
Neutron-capture-induced fission reactions also have significant importance in stellar
nucleosynthesis, being the end-point of the r-process~\cite{GOR13}. In this context, the systems that undergo fission are highly 
neutron-rich radioactive nuclei that are not accessible in laboratory settings. Consequently, a comprehensive theoretical description of 
the fission process across the nuclear chart is essential. While a complete microscopic description of fission remains beyond 
our current capabilities, the most reliable fission models are empirical or semi-empirical descriptions that require a substantial 
amount of experimental data from short-lived systems far from stability. Surrogate reactions, in which the nucleus of interest is produced through transfer- or gamma-induced reactions, offer an alternative to expand the range of accessible systems.

Surrogate reactions were initially employed to infer neutron-capture-induced fission cross sections~\cite{CRA70} 
under specific assumptions, such as the Weisskopf-Ewing limit. In this limit, the branching ratios of the compound 
nucleus are solely dependent on the excitation energy~\cite{Esch06,Esch12}:
\begin {equation}
\sigma_{n,x}(E_n) \approx \sigma_{CN}(E_n)\cdot P_{s,x}(E^{*}),
\end {equation}
where $\sigma_{n,x}(E_n)$ is the (n,x) reaction cross section at the neutron energy $E_n$, $\sigma_{CN}(E_n)$ is the total neutron-capture cross section, 
and $P_{s,x}(E^{*})$ is the branching ratio of the channel x. 
However, the Weisskopf-Ewing approximation has been proven to be invalid to obtain fission cross section 
for even-even fissioning systems. 
More realistic models of the spin and parity of states populated by surrogate reactions are required to extract reliably 
neutron-capture-induced fission cross sections~\cite{JuradoPRL}.

For the production of fission fragments, the comparison between neutron-capture-induced and surrogate-induced fission 
was extensively investigated and a consistent agreement was observed within 
the experimental uncertainties~\cite{Nishio03,Nishio16,ramPRC18}, suggesting that the initial spin-parity distribution 
of the fissioning system does not affect significantly the descent of the system from the fission barrier to the scission point. This observation was also supported 
by microscopic-level-densities models where no effect of the initial angular momentum is predicted in the fission yields for spins higher than J=0~\cite{Ward2017}.
Only recently, a dependence of symmetric fission on the reaction channel was observed in multinucleon-transfer-induced fission~\cite{Nishio20}; however, the effect of the damping of shell effects, the pre-saddle neutron-evaporation, and the excitation energy distribution of each particular transfer channel are mixed within the energy ranges investigated, hampering any comparison between reaction channels.

In the last decade, inverse kinematics has provided excellent opportunities to study surrogate reactions with measurements of complete fission-fragments distributions, fully identified in atomic and mass number with unprecedented resolutions~\cite{Sch00,caaPRC13,pell17,ramPRL,chat20}. However, in all these studies, the control over the excitation energy of the 
fissioning system was limited, 
either because the excitation energy of the system was not measured, or because the measured distribution of 
excitation energy was very broad. 

In this work, the total prompt neutron multiplicity 
of the fission of $^{240}$Pu is determined from the first measurement of post-neutron evaporation isotopic fission yields at various initial excitation energies 
via the surrogate 2p-transfer reaction $^{12}\text{C}(^{238}\text{U},^{240}\text{Pu}^*)^{10}\text{Be}$.
The total prompt neutron multiplicity of $^{238}$U, from inelastic-scattering-induced fission, is also presented for comparative purposes.

\begin{figure}[!]
\includegraphics[width=0.49\textwidth]{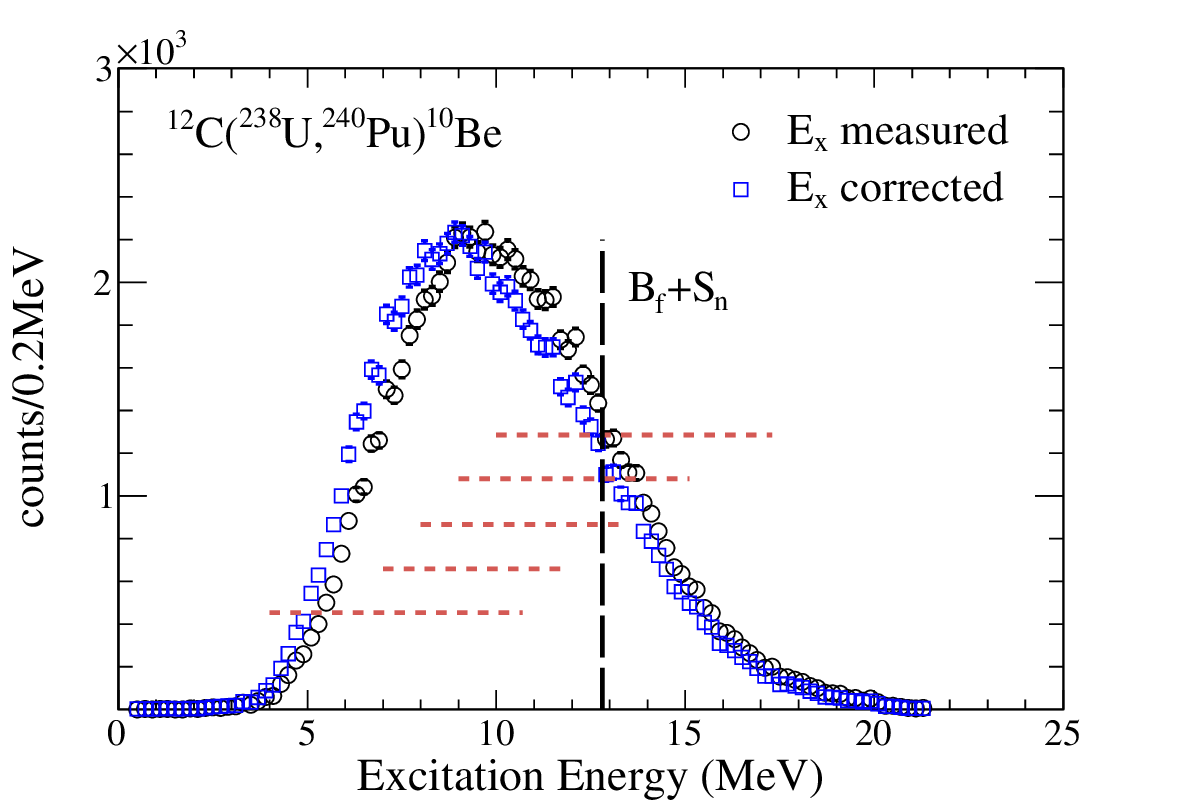}
\caption{Excitation energy distribution of $^{240}$Pu measured in coincidence with the detection of fission fragments (black circles)
and corrected for the probability of excitation of $^{10}$Be (blue squares). Dashed red lines represent the different excitation energy ranges selected for this analysis. The \textit{second-chance fission} energy threshold ($B_f + S_n$) is indicated by a vertical dashed black line.}
\label{fig:Ex}
\end{figure}

The measurement was conducted at GANIL, France, where a beam of $^{238}$U was accelerated to $6.14$~AMeV and 
impinged on a $100~\mu$g/cm$^2$-thick $^{12}$C target. The actinide $^{240}$Pu was produced through 2p-transfer reactions
$^{12}\text{C}(^{238}\text{U},^{240}\text{Pu}^*)^{10}\text{Be}$ in inverse kinematics, and underwent fission in flight. This fissioning system 
was identified by detecting the target-like recoil $^{10}$Be, as described in Ref.~\cite{rodPRC14}, and its excitation 
energy $E_x$ was determined event-by-event, ranging from $4$ to $20$~MeV. The distribution of $E_x$ measured in this 
experiment ($E_x^{m}$) is presented with black circles in Fig.~\ref{fig:Ex}. This distribution is corrected for the probability of populating excited states in the target-like recoil, which was measured to be $0.14 \pm 0.04$~\cite{rodPRC14}. 
The corrected $E_x$ distribution is presented with blue squares in Fig.~\ref{fig:Ex}. A black dashed line in the same figure indicates the threshold excitation energy of the \textit{second-chance fission}. This is the minimum excitation energy that the fissioning system needs to evaporate one neutron ($S_n$) and overcome the fission barrier ($B_f$) with the remaining energy. In the same experiment, inelastic-scattering-induced fission of $^{238}$U was also measured by detecting the target-like recoil $^{12}$C.

The selection of $^{10}$Be events includes a $5\pm 2$~\% of contamination due to an overlap with $^{9}$Be, while the selection of $^{12}$C includes a $3\pm 2$~\% of contamination from $^{13}$C and $^{14}$C~\cite{rodPRC14}. The subtraction of this contamination results in a maximum shift of 0.04~MeV towards lower $E_x$. This value is two orders of magnitude smaller than the experimental sensitivity and it is therefore neglected in the calculation of $E_x$.

When decaying in flight, the fissioning systems split into two fragments emitted within a cone of $30$ degrees in the laboratory frame.
For each fission event, one of the fission fragments was detected at the focal plane setup of the magnetic spectrometer VAMOS++ and it was isotopically identified~\cite{rejNIM,LemNIM}, if it was within the spectrometer acceptance. 

\begin{table} [!t]
\begin{center}
  \caption{\label{table1}Characteristics of the excitation energy ranges of $^{240}$Pu and $^{238}$U selected for the analysis are presented. 
These ranges encompass the limits of each range, along with the average values before and after correction for the target-like recoil excitation. 
Additionally, standard deviations are provided for each range.}
  \begin{tabular}{|c|c|c|c|}
    \hline\hline
    ~$E_{x}$ limits~  &~Average $E_{x}^{m}$~&~Average~$E_{x}$~&~$SD_{E_{x}}$~\\
   (MeV) & (MeV) & (MeV) & (MeV) \\\hline
 \multicolumn{4}{|c|}{$^{240}$Pu} \\\hline
4.0 - 10.7 & 8.5 & 8.2 & 1.46\\
7.0 - 11.8 & 9.5 & 9.0 & 1.32\\
8.0 - 13.3 & 10.5 & 10.0 & 1.46\\
9.0 - 15.1& 11.5 & 10.9 & 1.63\\
10.0 - 17.3 & 12.5 & 11.9 & 1.79\\\hline
 \multicolumn{4}{|c|}{$^{238}$U} \\\hline
4.5 - 9.5 & 6.7 & 6.6 & 1.33\\
5.5 - 12.0 & 7.7 & 7.5 & 1.66\\
6.0 - 17.5 & 8.7 & 8.6 & 2.50\\
6.8 - 20.0 & 9.7 & 8.9 & 2.86\\\hline
  \end{tabular}
\end{center}
\end{table}

The isotopic fission yields of 
$^{240}$Pu were obtained, as described in Ref~\cite{ramPRC18}, for different ranges of $E_x$ shown with red dashed lines in Fig.~\ref{fig:Ex}. 
They were determined requiring $4\times10^4$ events per steps of $1$~MeV of the average 
$E_x^{m}$. A similar procedure was applied to $^{238}$U. The limits, average values, and standard deviation of each $E_x$ range are detailed in 
Table~\ref{table1}.

\begin{figure}[!]
\includegraphics[width=0.49\textwidth]{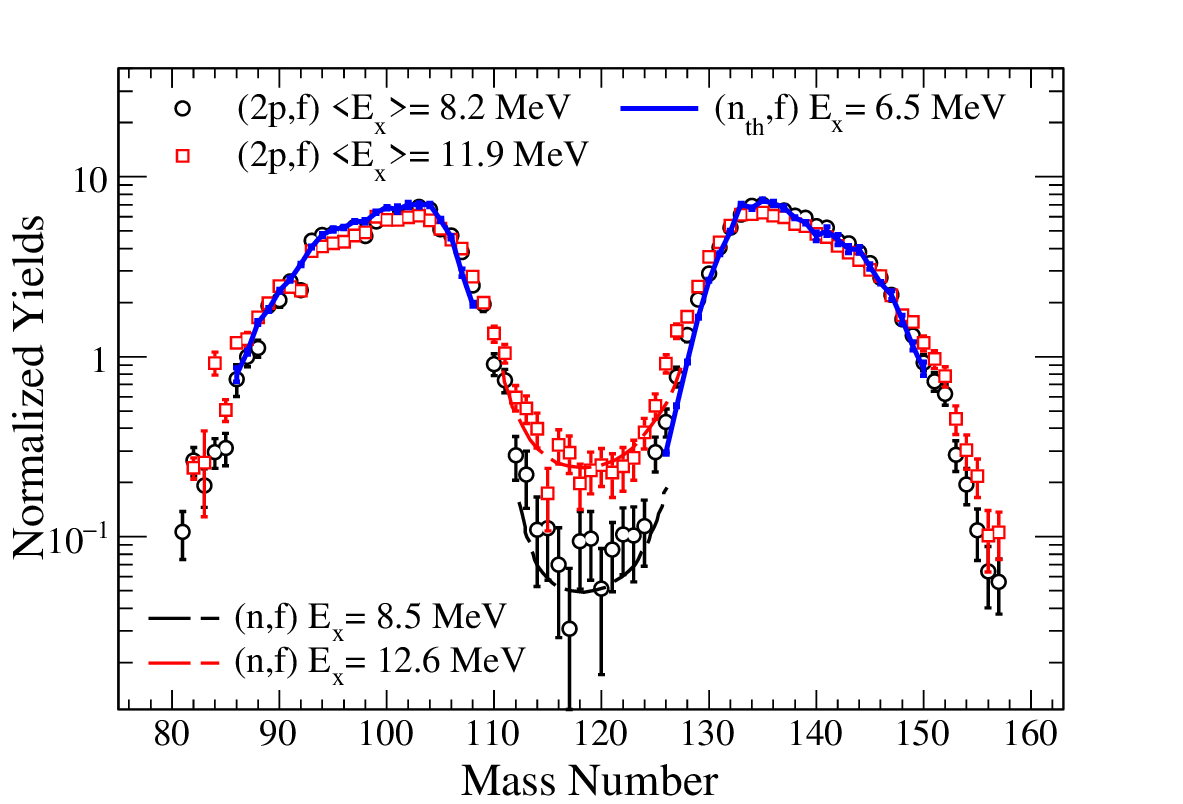}
\caption{The distribution of fission yields of $^{240}$Pu as a function of the post neutron-evaporation fission-fragment mass is plotted for two average excitation energies: $8.2$~MeV (black circles) and $11.9$~MeV (red squares). The data is compared to neutron-capture-induced fission at $6.5$~MeV from Refs.~\cite{schnpa,BaiPRC}, and to extrapolated data for the valley region at excitation energies of $8.5$~MeV and $12.6$~MeV, as presented in Ref.~\cite{GinPRC}.}
\label{fig:YA}
\end{figure}

The evolution of the fission fragment yields as a function of the initial $E_x$ of the system 
is presented in Figure~\ref{fig:YA}. The fission yields are presented as a function of the post-neutron evaporation
fission fragment mass at two average excitation energies: $8.2$~MeV (black circles) and $11.9$~MeV
(red squares). The uncertainties of the present fission yields are determined as the quadratic sum of statistical and systematic sources. Systematic uncertainties range from 4\% to 10\%, taking into account normalization, intrinsic efficiency, and spectrometer acceptance uncertainties. 

As $E_x$ increases, it is observed that the population of the symmetry region is also increasing. This is compared with previous measurements 
of neutron-capture-induced fission of $^{239}$Pu~\cite{GinPRC}, which have been extrapolated from cumulative fission 
yields at two neutron energies equivalent to $8.5$~MeV and $12.6$~MeV of $E_x$, depicted as black and red dashed lines, respectively. 
Within uncertainties, both datasets show a similar damping of shell effects.

The mass asymmetry region of the present data is also compared in Fig.~\ref{fig:YA} with thermal-neutron-capture-induced fission with $E_x = 6.5$~MeV from Refs.~\cite{schnpa,BaiPRC}, represented with blue lines. There is a uniform 
evolution of the yields for the most produced fragments: the higher $E_x$, the lower the yields become, compensating the population of the symmetric region. A coherent evolution of the asymmetric components of fission is observed 
between both sets of data.

\begin{figure}[!]
\includegraphics[width=0.49\textwidth]{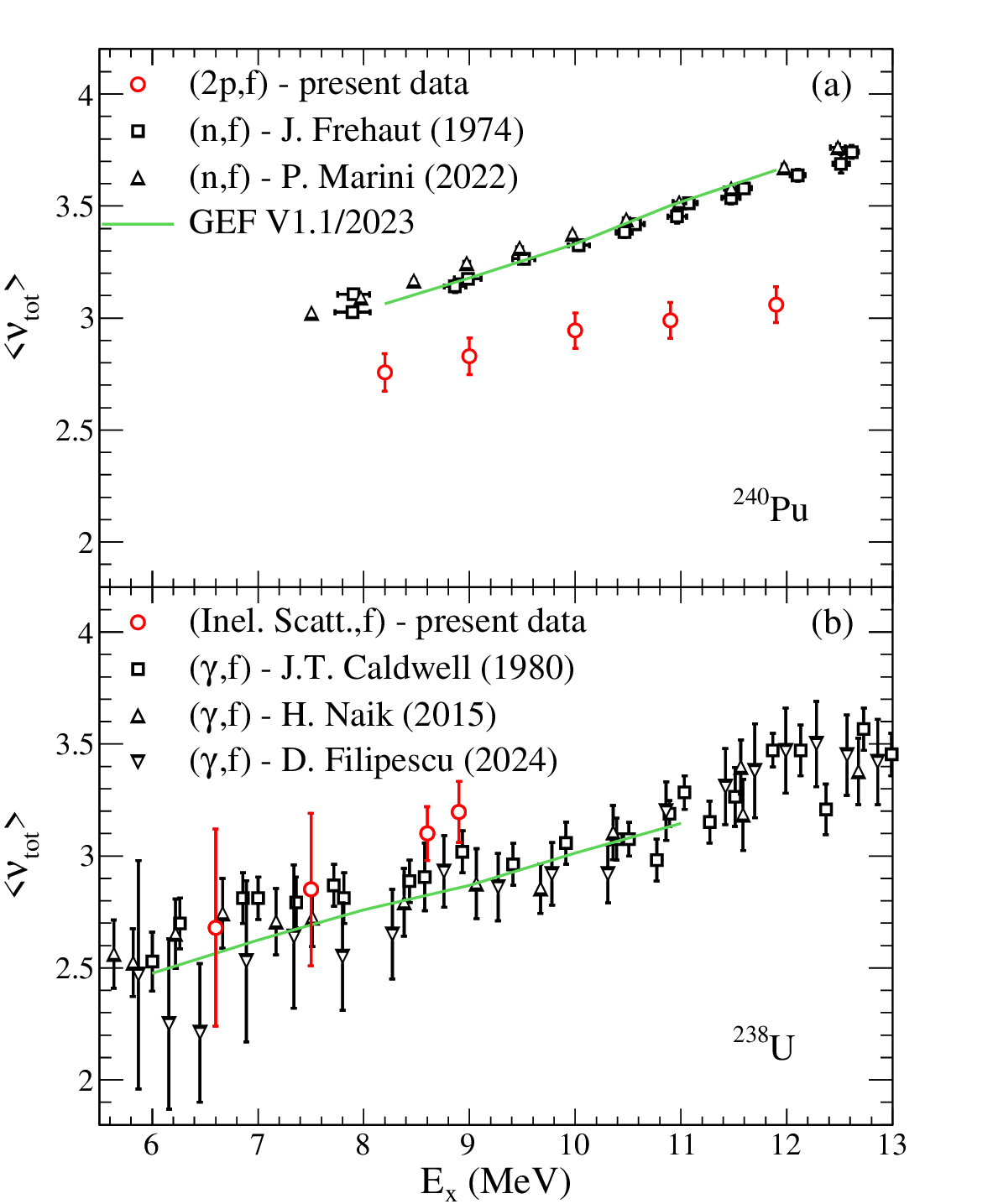}
\caption{(a) Average total prompt neutron multiplicity of 2-proton transfer-induced fission of $^{240}$Pu as a function of the initial excitation energy of the system. The present data (red circles) is compared to neutron-capture-induced fission on $^{239}$Pu from Refs.~\cite{Fre80,Mar22} (black symbols), and to GEF calculations (green line). 
(b) Average total prompt neutron multiplicity of inelastic-scattering-induced fission of $^{238}$U as a function of 
the initial excitation energy of the system. The present data (red circles) is compared to gamma-induced fission of $^{238}$U from Refs.~\cite{Caldwell80,Naik15,Filipescu24} (black symbols) and to GEF (green line).}
\label{fig:Nevap}
\end{figure}

The prompt neutron multiplicity ---i.e., the number of neutrons evaporated in the fission process before the radioactive decay of fission products--- can be retreived by summing the neutron content of both post-neutron-evaporation fission fragments and comparing it to the neutron content of the fissioning system.
The isotopic identification of the complete distribution of post-neutron-evaporation fission fragments enables the calculation of the average prompt neutron multiplicity as
\begin{equation}
\label{eq_nu}
\langle \nu_{tot}\rangle = N_{fis}-\frac{1}{100}\sum_Z^{Z_{fis}/2}{(\langle N\rangle|_Z +\langle N\rangle|_{Z_{fis}-Z})\times Y(Z) } ,
\end{equation}
where
\begin{equation}
\langle N\rangle|_Z = \frac{1}{Y(Z)}\sum_N{N(Z)\times Y(N,Z) } 
\end{equation}
is the average number of neutrons of each elemental fragment $Z$,
$Y(N,Z)$ is the yield of the isotope $(N,Z)$, $Y(Z)=\sum_N{Y(N,Z)}$ is the yield of the element $Z$, $Z_{fis}=94$ is 
the number of protons of the compound nucleus, and $N_{fis}=146.05\pm0.02$ is the number of neutrons of the compound nucleus. The latter takes into account a 5\% contamination from $^{241}$Pu.

The resulting total neutron multiplicity ($\langle \nu_{tot}\rangle$) as a function of the average $E_x$ of $^{240}$Pu is presented in Fig.~\ref{fig:Nevap}(a) with red circles. The asociated uncertainties take into account the uncertainty of the isotopic fission yields and the uncertainty in the determination of the fissioning system. An additional uncertainty of 0.1\% is included in the neutron multiplicity to account for the 2\% maximum difference in neutron evaporation between $^{241}$Pu and $^{240}$Pu, as determined using the General Description of Fission Observables model (GEF)~\cite{schGEF}, a state-of-the-art semi-empirical fission model currently incorporated into various evaluation data bases, such as JEFF-3.3.

The present data is compared with the total prompt neutron multiplicity measured in 
neutron-capture-induced fission on $^{239}$Pu(n,f) from Refs.~\cite{Fre80,Mar22} (black squares and triangles, respectively). 
The present data is also compared with GEF (green line). GEF accurately reproduces the neutron-induced data sets while failing to reproduce the present one.
The prompt neutron multiplicity from the surrogate 2-proton-transfer reaction 
$^{12}\text{C}(^{238}\text{U},^{240}\text{Pu}^*)^{10}\text{Be}$, is systematically lower than this in neutron-capture-induced 
fission at the same initial $E_x$, with a difference that increases with increasing $E_x$. 
This discrepancy presents a clear challenge in describing multi-nucleon transfer as a surrogate reaction for 
neutron-capture-induced fission. 

The average total prompt neutron multiplicity of $^{238}$U is presented in Fig.~\ref{fig:Nevap}(b) with red circles as a function of 
the average initial $E_x$. The present data is compared with the total prompt neutron multiplicity 
measured in $\gamma$-induced fission of $^{238}\text{U}(\gamma,f)$ from Refs.~\cite{Caldwell80,Naik15,Filipescu24} 
(black symbols) and with GEF (green line). In this case, both, inelastic-scattering-induced and gamma-induced 
fission exhibit similar prompt neutron evaporation. GEF also reproduces the overall trend, but it locally underestimates the neutron evaporation around $E_x\sim$9~MeV.

The observed difference in the prompt-neutron multiplicity for $^{240}$Pu can be attributed to the varying angular momenta 
populated by the different incoming channels. Two-proton transfer reactions are expected to induce a higher angular momentum into 
the fissioning system compared to neutron-capture reactions. A calculation with the code TALYS 2.0~\cite{TALYS} predicts an angular momentum of $3~\hbar$ for neutron-capture reactions at the present $E_x$. However,the angular momentum generated in the fissioning system by 2p-transfer reactions
is estimated to be of the order of $10~\hbar$. As an example, the spin population in 2-p transfer reactions involving a $^{18}$O beam impinging on a $^{237}$Np target at Coulomb energies was experimentally measured at $10.3\pm2.0\hbar$~\cite{Tanaka22}.
This angular momentum is expected to influence the competition between neutron evaporation and $\gamma$-ray emission 
along the fission path, as higher angular momentum increases the probability of $\gamma$-ray emission while reducing the probability of neutron evaporation.

On the contrary, for $^{238}$U, both $\gamma$-ray absorption and inelastic scattering are expected to populate 
comparable angular momentum distributions. 
In the present data, inelastic scattering reactions populate an average angular momentum of $2.4~\hbar$.
This was estimated by using three EXOGAM detectors~\cite{EXOGAM}, positioned at 
backward angles at $15$~cm from the target, for measuring the $\gamma$-ray decay of the excited states of $^{238}$U in 
competition with fission events. In the case of $^{240}$Pu, this measurement was not accesible, as most of the events undergo 
fission at the studied $E_x$ range, and the corresponding $\gamma$-ray decay rate was too weak. 

\begin{figure}[!]
\includegraphics[width=0.49\textwidth]{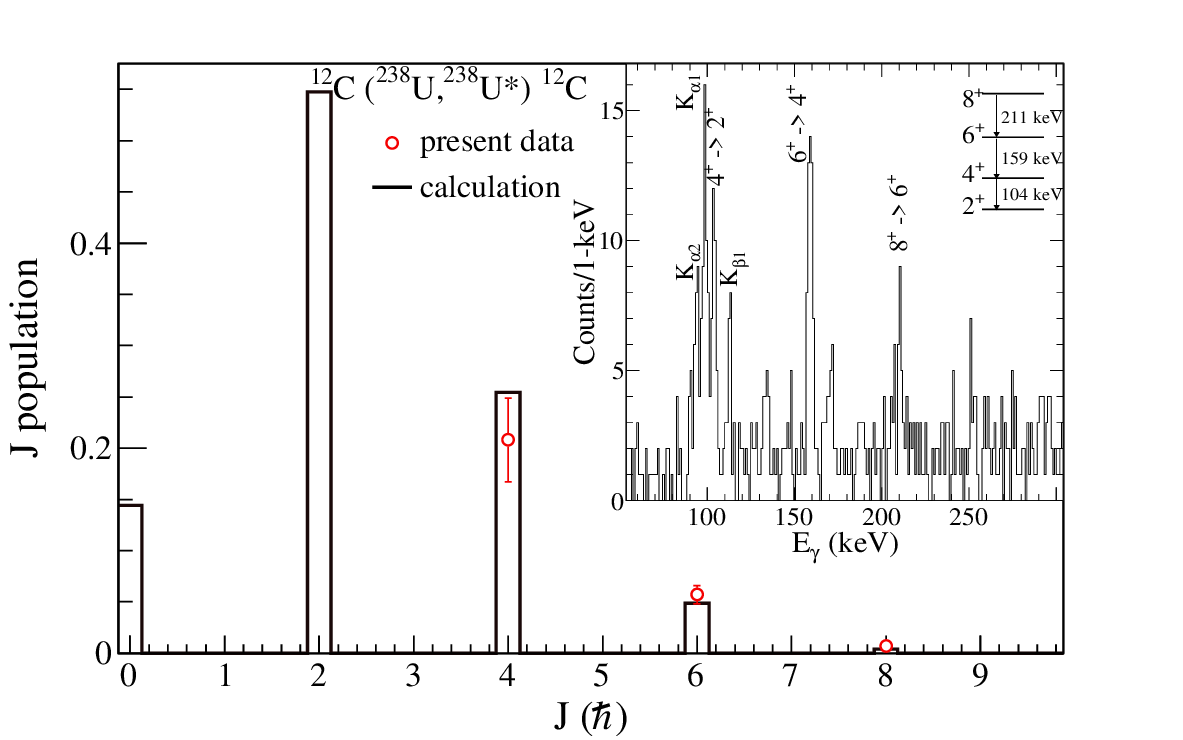}
\caption{Population of spin $J$ in $^{238}\text{U}^*$ excited through inelastic scattering reactions
$^{12}\text{C}(^{238}\text{U},^{238}\text{U}^*)^{12}\text{C}$ with an energy $8~\%$ above the Coulomb barrier
Present data (red circles) is compared with a calculation of Coulomb excitation~\cite{GOSIA} (back bars).
An average angular momentum of $2.4~\hbar$ was determined. 
(Inset) Gamma decay spectrum of $^{238}\text{U}^*$ in the present data. }
\label{fig:GammaU}
\end{figure}

Figure~\ref{fig:GammaU} presents the population of spin in $^{238}\text{U}^*$ excited through inelastic scattering 
$^{12}\text{C}(^{238}\text{U},^{238}\text{U}^*)^{12}\text{C}$. The spin distribution is determined by measuring 
the $\gamma$-ray cascade emitted by the ground-state rotational band  of $^{238}\text{U}^*$. In the inset, the 
$8^+\rightarrow 6^+$, $6^+\rightarrow 4^+$, and $4^+\rightarrow 2^+$ transitions are clearly identified
along with the $\text{K}_{\alpha1}$, $\text{K}_{\alpha2}$, and $\text{K}_{\beta1}$ X-rays. The spin distribution 
was reconstructed by correcting the intensity of each transition for the feeding of the higher-lying states, 
the detection efficiency, and the internal conversion probability. The experimental result (red circles) is compared with 
a realistic calculation of Coulomb excitation~\cite{GOSIA} (back bars) from which the average spin population was determined. 

In order to investigate the origin of the difference in prompt neutron multiplicity of $^{240}$Pu between 2p-transfer-induced and neutron-capture-induced fission, prompt neutron multiplicity from present data was reproduced by fitting the de-excitation of fully accelerated fission fragments with the FIFRELIN code~\cite{fifrelin}.
 Primary fission fragments obtained from the GEF code were used as an input for the FIFRELIN calculation and the de-excitation was computed as a statistical process following 
the Hauser-Feshbach formalism with explicit competition between neutrons and $\gamma$-rays based on level densities calculated from a Hartree-Fock-Bogolyubov model. As a result of the fit, the primary fission fragments from 
2p-transfer reactions must have an average spin population between $15~\hbar$ and $18~\hbar$, 
in contrast to $7~\hbar$ to $8~\hbar$ in neutron-capture-induced fission. The latter was determined by fitting 
the prompt neutron multiplicity obtained from the JEFF-3 nuclear database~\cite{JEFF} using a similar procedure.


The average spin values of neutron-capture-induced fission fragments estimated from this fit are consistent with experimental data, where fragments produced in spontaneous fission of $^{252}$Cf  and neutron-capture-induced fission of $^{238}$U were observed to follow the same spin population~\cite{WilsonNature}. However, the average spin values fitted with the same procedure to the present data are too high to be compatible even with excited states measured in fragments produced in fusion-induced fission reactions, where the initial spin is of the order of $20~\hbar$~\cite{Shrivastava09,Navin14}. Thus, fragments with spin populations of $15~\hbar$ to $18~\hbar$ are also incompatible with transfer-induced reactions with an initial spin of the order of $10~\hbar$.

Therefore, part of the angular momentum excess induced in the surrogate transfer reaction must be depleted before the scission point. Consequently, the reduction of the prompt neutron multiplicity from 2p-transfer compared to neutron-capture-induced fission can be attributed, at least partially, to a reduction in the pre-scission neutron evaporation. Pre-scission neutron emission in low-excitation compound nuclei has long been a subject of investigation in fission, with calculations suggesting that more than 10\% of the total neutrons are emitted before scission~\cite{carjan07,capote16,Abdurrahman24}.

This observation is also consistent and complementary to recent measurements of fission-fragment isomeric yields ratios that suggest that at least 40\% of the difference in the initial angular momentum between surrogate-induced and neutron-capture-induced fission contributes to the spin of the fragments~\cite{cannarozzo24}.  

The fission barrier of $^{240}$Pu extracted from 2p-transfer was experimentally observed not to differ from the one extracted from neutron-capture induced fission~\cite{rodPRC14}. Hence, in the scenario where $E_x$ is less than $12.8$~MeV, neutron evaporation before saddle is energetically infeasible~\cite{JuradoPRL}. In addition, the good agreement in the relative fission-fragment production at the symmetry between present data and neutron-induced fission indicates that the system reaches the saddle point with similar excitation energy in both reactions. In these conditions, the neutron multiplicity deficit certainly arises from a reduction of the neutron evaporation of the system between the saddle and the scission point and, possibly, from a reduction of the neutron evaporation of primary fission fragments.


In summary, the present data provides the first experimental evidence of a reduction of the prompt neutron evaporation in a surrogate reaction for neutron-induced fission. The suppression of neutron evaporation in the transfer-induced fission reaction $^{12}\text{C}(^{238}\text{U},^{240}\text{Pu}^*)^{10}\text{Be}$ compared to the neutron-induced reaction $^{239}$Pu(n,f) is interpreted as the effect of the additional angular momentum populated in the fissioning system.
This observation presents a challenge for the application of surrogate reactions as an equivalent of neutron-capture-induced fission, for instance, in criticality and heat calculations of next-generation reactors utilizing surrogate data, as in their breeding capability, with prompt-neutron multiplicity being the primary factor in sustaining the fission chain~\cite{Duderstadt,Marguet2017,Doligez}. 
Besides the effect on neutron evaporation, fission fragment yields are found to be robust against angular momentum variations, suggesting that the shape of the potential energy surface of the fission process is rather insensitive to moderate variations of the initial angular momentum of the fissioning system.

A comprehensive theoretical study of the interplay between the reaction mechanism of multi-nucleon transfer 
of heavy nuclei and fission dynamics is necessary to better understand the limits of surrogate methods. Further, combined measurements of both neutron evaporation and $\gamma$-ray emission, 
are required to quantify the impact of angular momentum on their competition within the fission process. The acquisition of higher precision experimental fission-yield data is planned 
to investigate the limits of the surrogate approach for fission.

\begin{acknowledgments}
This work was partially supported by the Spanish Ministry of Research and Innovation under the budget items FPA2010- 22174-C02-01 and RYC-2012-11585. The excellent support from the GANIL staff during the experiment is acknowledged.
\end{acknowledgments}

\Urlmuskip=0mu plus 1mu\relax
\bibliography{Paper_Pu_Ex_6_ReSumArxiv} 

\begin{thebibliography}{44}%
\makeatletter
\providecommand \@ifxundefined [1]{%
 \@ifx{#1\undefined}
}%
\providecommand \@ifnum [1]{%
 \ifnum #1\expandafter \@firstoftwo
 \else \expandafter \@secondoftwo
 \fi
}%
\providecommand \@ifx [1]{%
 \ifx #1\expandafter \@firstoftwo
 \else \expandafter \@secondoftwo
 \fi
}%
\providecommand \natexlab [1]{#1}%
\providecommand \enquote  [1]{``#1''}%
\providecommand \bibnamefont  [1]{#1}%
\providecommand \bibfnamefont [1]{#1}%
\providecommand \citenamefont [1]{#1}%
\providecommand \href@noop [0]{\@secondoftwo}%
\providecommand \href [0]{\begingroup \@sanitize@url \@href}%
\providecommand \@href[1]{\@@startlink{#1}\@@href}%
\providecommand \@@href[1]{\endgroup#1\@@endlink}%
\providecommand \@sanitize@url [0]{\catcode `\\12\catcode `\$12\catcode
  `\&12\catcode `\#12\catcode `\^12\catcode `\_12\catcode `\%12\relax}%
\providecommand \@@startlink[1]{}%
\providecommand \@@endlink[0]{}%
\providecommand \url  [0]{\begingroup\@sanitize@url \@url }%
\providecommand \@url [1]{\endgroup\@href {#1}{\urlprefix }}%
\providecommand \urlprefix  [0]{URL }%
\providecommand \Eprint [0]{\href }%
\providecommand \doibase [0]{https://doi.org/}%
\providecommand \selectlanguage [0]{\@gobble}%
\providecommand \bibinfo  [0]{\@secondoftwo}%
\providecommand \bibfield  [0]{\@secondoftwo}%
\providecommand \translation [1]{[#1]}%
\providecommand \BibitemOpen [0]{}%
\providecommand \bibitemStop [0]{}%
\providecommand \bibitemNoStop [0]{.\EOS\space}%
\providecommand \EOS [0]{\spacefactor3000\relax}%
\providecommand \BibitemShut  [1]{\csname bibitem#1\endcsname}%
\let\auto@bib@innerbib\@empty
\bibitem [{NEA(2022)}]{NEA}%
  \BibitemOpen
  \href
  {https://www.oecd-nea.org/upload/docs/application/pdf/2023-12/nea_ar_2022.pdf}
  {\bibinfo {title} {{2022 OECD Nuclear Energy Agency Annual Report, Report No.
  7653}}} (\bibinfo {year} {2022})\BibitemShut {NoStop}%
\bibitem [{\citenamefont {Goriely}\ \emph {et~al.}(2013)\citenamefont
  {Goriely}, \citenamefont {Sida}, \citenamefont {Lema{\^{i}}tre},
  \citenamefont {Panebianco}, \citenamefont {Dubray}, \citenamefont {Hilaire},
  \citenamefont {Bauswein},\ and\ \citenamefont {Janka}}]{GOR13}%
  \BibitemOpen
  \bibfield  {author} {\bibinfo {author} {\bibfnamefont {S.}~\bibnamefont
  {Goriely}}, \bibinfo {author} {\bibfnamefont {J.~L.}\ \bibnamefont {Sida}},
  \bibinfo {author} {\bibfnamefont {J.~F.}\ \bibnamefont {Lema{\^{i}}tre}},
  \bibinfo {author} {\bibfnamefont {S.}~\bibnamefont {Panebianco}}, \bibinfo
  {author} {\bibfnamefont {N.}~\bibnamefont {Dubray}}, \bibinfo {author}
  {\bibfnamefont {S.}~\bibnamefont {Hilaire}}, \bibinfo {author} {\bibfnamefont
  {A.}~\bibnamefont {Bauswein}},\ and\ \bibinfo {author} {\bibfnamefont
  {H.~T.}\ \bibnamefont {Janka}},\ }\href
  {https://doi.org/10.1103/PhysRevLett.111.242502} {\bibfield  {journal}
  {\bibinfo  {journal} {Phys. Rev. Lett.}\ }\textbf {\bibinfo {volume} {111}},\
  \bibinfo {pages} {242502} (\bibinfo {year} {2013})}\BibitemShut {NoStop}%
\bibitem [{\citenamefont {Cramer}\ and\ \citenamefont {Britt}(1970)}]{CRA70}%
  \BibitemOpen
  \bibfield  {author} {\bibinfo {author} {\bibfnamefont {J.~D.}\ \bibnamefont
  {Cramer}}\ and\ \bibinfo {author} {\bibfnamefont {H.~C.}\ \bibnamefont
  {Britt}},\ }\href {https://doi.org/10.1103/PhysRevC.2.2350} {\bibfield
  {journal} {\bibinfo  {journal} {Phys. Rev. C}\ }\textbf {\bibinfo {volume}
  {2}},\ \bibinfo {pages} {2350} (\bibinfo {year} {1970})}\BibitemShut
  {NoStop}%
\bibitem [{\citenamefont {Escher}\ and\ \citenamefont
  {Dietrich}(2006)}]{Esch06}%
  \BibitemOpen
  \bibfield  {author} {\bibinfo {author} {\bibfnamefont {J.~E.}\ \bibnamefont
  {Escher}}\ and\ \bibinfo {author} {\bibfnamefont {F.~S.}\ \bibnamefont
  {Dietrich}},\ }\href {https://doi.org/10.1103/PhysRevC.74.054601} {\bibfield
  {journal} {\bibinfo  {journal} {Phys. Rev. C}\ }\textbf {\bibinfo {volume}
  {74}},\ \bibinfo {pages} {054601} (\bibinfo {year} {2006})}\BibitemShut
  {NoStop}%
\bibitem [{\citenamefont {Escher}\ \emph {et~al.}(2012)\citenamefont {Escher},
  \citenamefont {Burke}, \citenamefont {Dietrich}, \citenamefont {Scielzo},
  \citenamefont {Thompson},\ and\ \citenamefont {Younes}}]{Esch12}%
  \BibitemOpen
  \bibfield  {author} {\bibinfo {author} {\bibfnamefont {J.~E.}\ \bibnamefont
  {Escher}}, \bibinfo {author} {\bibfnamefont {J.~T.}\ \bibnamefont {Burke}},
  \bibinfo {author} {\bibfnamefont {F.~S.}\ \bibnamefont {Dietrich}}, \bibinfo
  {author} {\bibfnamefont {N.~D.}\ \bibnamefont {Scielzo}}, \bibinfo {author}
  {\bibfnamefont {I.~J.}\ \bibnamefont {Thompson}},\ and\ \bibinfo {author}
  {\bibfnamefont {W.}~\bibnamefont {Younes}},\ }\href
  {https://doi.org/10.1103/RevModPhys.84.353} {\bibfield  {journal} {\bibinfo
  {journal} {Rev. Mod. Phys.}\ }\textbf {\bibinfo {volume} {84}},\ \bibinfo
  {pages} {353} (\bibinfo {year} {2012})}\BibitemShut {NoStop}%
\bibitem [{\citenamefont {P{\'e}rez~S{\'a}nchez}\ \emph
  {et~al.}(2020)\citenamefont {P{\'e}rez~S{\'a}nchez}, \citenamefont {Jurado},
  \citenamefont {M{\'e}ot}, \citenamefont {Roig}, \citenamefont {Dupuis},
  \citenamefont {Bouland}, \citenamefont {Denis-Petit}, \citenamefont {Marini},
  \citenamefont {Mathieu}, \citenamefont {Tsekhanovich}, \citenamefont
  {A{\"i}che}, \citenamefont {Audouin}, \citenamefont {Cannes}, \citenamefont
  {Czajkowski}, \citenamefont {Delpech}, \citenamefont {G{\"o}rgen},
  \citenamefont {Guttormsen}, \citenamefont {Henriques}, \citenamefont
  {Kessedjian}, \citenamefont {Nishio}, \citenamefont {Ramos}, \citenamefont
  {Siem},\ and\ \citenamefont {Zeiser}}]{JuradoPRL}%
  \BibitemOpen
  \bibfield  {author} {\bibinfo {author} {\bibfnamefont {R.}~\bibnamefont
  {P{\'e}rez~S{\'a}nchez}}, \bibinfo {author} {\bibfnamefont {B.}~\bibnamefont
  {Jurado}}, \bibinfo {author} {\bibfnamefont {V.}~\bibnamefont {M{\'e}ot}},
  \bibinfo {author} {\bibfnamefont {O.}~\bibnamefont {Roig}}, \bibinfo {author}
  {\bibfnamefont {M.}~\bibnamefont {Dupuis}}, \bibinfo {author} {\bibfnamefont
  {O.}~\bibnamefont {Bouland}}, \bibinfo {author} {\bibfnamefont
  {D.}~\bibnamefont {Denis-Petit}}, \bibinfo {author} {\bibfnamefont
  {P.}~\bibnamefont {Marini}}, \bibinfo {author} {\bibfnamefont
  {L.}~\bibnamefont {Mathieu}}, \bibinfo {author} {\bibfnamefont
  {I.}~\bibnamefont {Tsekhanovich}}, \bibinfo {author} {\bibfnamefont
  {M.}~\bibnamefont {A{\"i}che}}, \bibinfo {author} {\bibfnamefont
  {L.}~\bibnamefont {Audouin}}, \bibinfo {author} {\bibfnamefont
  {C.}~\bibnamefont {Cannes}}, \bibinfo {author} {\bibfnamefont
  {S.}~\bibnamefont {Czajkowski}}, \bibinfo {author} {\bibfnamefont
  {S.}~\bibnamefont {Delpech}}, \bibinfo {author} {\bibfnamefont
  {A.}~\bibnamefont {G{\"o}rgen}}, \bibinfo {author} {\bibfnamefont
  {M.}~\bibnamefont {Guttormsen}}, \bibinfo {author} {\bibfnamefont
  {A.}~\bibnamefont {Henriques}}, \bibinfo {author} {\bibfnamefont
  {G.}~\bibnamefont {Kessedjian}}, \bibinfo {author} {\bibfnamefont
  {K.}~\bibnamefont {Nishio}}, \bibinfo {author} {\bibfnamefont
  {D.}~\bibnamefont {Ramos}}, \bibinfo {author} {\bibfnamefont
  {S.}~\bibnamefont {Siem}},\ and\ \bibinfo {author} {\bibfnamefont
  {F.}~\bibnamefont {Zeiser}},\ }\href
  {https://doi.org/10.1103/PhysRevLett.125.122502} {\bibfield  {journal}
  {\bibinfo  {journal} {Phys. Rev. Lett.}\ }\textbf {\bibinfo {volume} {125}},\
  \bibinfo {pages} {122502} (\bibinfo {year} {2020})}\BibitemShut {NoStop}%
\bibitem [{\citenamefont {Nishio}\ \emph {et~al.}(2003)\citenamefont {Nishio},
  \citenamefont {Ikezoe}, \citenamefont {Nagame}, \citenamefont {Mitsuoka},
  \citenamefont {Nishinaka}, \citenamefont {Duan}, \citenamefont {Satou},
  \citenamefont {Goto}, \citenamefont {Asai}, \citenamefont {Haba},
  \citenamefont {Tsukada}, \citenamefont {Shinohara}, \citenamefont
  {Ichikawa},\ and\ \citenamefont {Ohsawa}}]{Nishio03}%
  \BibitemOpen
  \bibfield  {author} {\bibinfo {author} {\bibfnamefont {K.}~\bibnamefont
  {Nishio}}, \bibinfo {author} {\bibfnamefont {H.}~\bibnamefont {Ikezoe}},
  \bibinfo {author} {\bibfnamefont {Y.}~\bibnamefont {Nagame}}, \bibinfo
  {author} {\bibfnamefont {S.}~\bibnamefont {Mitsuoka}}, \bibinfo {author}
  {\bibfnamefont {I.}~\bibnamefont {Nishinaka}}, \bibinfo {author}
  {\bibfnamefont {L.}~\bibnamefont {Duan}}, \bibinfo {author} {\bibfnamefont
  {K.}~\bibnamefont {Satou}}, \bibinfo {author} {\bibfnamefont
  {S.}~\bibnamefont {Goto}}, \bibinfo {author} {\bibfnamefont {M.}~\bibnamefont
  {Asai}}, \bibinfo {author} {\bibfnamefont {H.}~\bibnamefont {Haba}}, \bibinfo
  {author} {\bibfnamefont {K.}~\bibnamefont {Tsukada}}, \bibinfo {author}
  {\bibfnamefont {N.}~\bibnamefont {Shinohara}}, \bibinfo {author}
  {\bibfnamefont {S.}~\bibnamefont {Ichikawa}},\ and\ \bibinfo {author}
  {\bibfnamefont {T.}~\bibnamefont {Ohsawa}},\ }\href
  {https://doi.org/10.1103/PhysRevC.67.014604} {\bibfield  {journal} {\bibinfo
  {journal} {Phys. Rev. C}\ }\textbf {\bibinfo {volume} {67}},\ \bibinfo
  {pages} {014604} (\bibinfo {year} {2003})}\BibitemShut {NoStop}%
\bibitem [{\citenamefont {L{\'e}guillon}\ \emph {et~al.}(2016)\citenamefont
  {L{\'e}guillon}, \citenamefont {Nishio}, \citenamefont {Hirose},
  \citenamefont {Makii}, \citenamefont {Nishinaka}, \citenamefont {Orlandi},
  \citenamefont {Tsukada}, \citenamefont {Smallcombe}, \citenamefont {Chiba},
  \citenamefont {Aritomo}, \citenamefont {Ohtsuki}, \citenamefont {Tatsuzawa},
  \citenamefont {Takaki}, \citenamefont {Tamura}, \citenamefont {Goto},
  \citenamefont {Tsekhanovich}, \citenamefont {Petrache},\ and\ \citenamefont
  {Andreyev}}]{Nishio16}%
  \BibitemOpen
  \bibfield  {author} {\bibinfo {author} {\bibfnamefont {R.}~\bibnamefont
  {L{\'e}guillon}}, \bibinfo {author} {\bibfnamefont {K.}~\bibnamefont
  {Nishio}}, \bibinfo {author} {\bibfnamefont {K.}~\bibnamefont {Hirose}},
  \bibinfo {author} {\bibfnamefont {H.}~\bibnamefont {Makii}}, \bibinfo
  {author} {\bibfnamefont {I.}~\bibnamefont {Nishinaka}}, \bibinfo {author}
  {\bibfnamefont {R.}~\bibnamefont {Orlandi}}, \bibinfo {author} {\bibfnamefont
  {K.}~\bibnamefont {Tsukada}}, \bibinfo {author} {\bibfnamefont
  {J.}~\bibnamefont {Smallcombe}}, \bibinfo {author} {\bibfnamefont
  {S.}~\bibnamefont {Chiba}}, \bibinfo {author} {\bibfnamefont
  {Y.}~\bibnamefont {Aritomo}}, \bibinfo {author} {\bibfnamefont
  {T.}~\bibnamefont {Ohtsuki}}, \bibinfo {author} {\bibfnamefont
  {R.}~\bibnamefont {Tatsuzawa}}, \bibinfo {author} {\bibfnamefont
  {N.}~\bibnamefont {Takaki}}, \bibinfo {author} {\bibfnamefont
  {N.}~\bibnamefont {Tamura}}, \bibinfo {author} {\bibfnamefont
  {S.}~\bibnamefont {Goto}}, \bibinfo {author} {\bibfnamefont {I.}~\bibnamefont
  {Tsekhanovich}}, \bibinfo {author} {\bibfnamefont {C.}~\bibnamefont
  {Petrache}},\ and\ \bibinfo {author} {\bibfnamefont {A.}~\bibnamefont
  {Andreyev}},\ }\href
  {https://doi.org/https://doi.org/10.1016/j.physletb.2016.08.010} {\bibfield
  {journal} {\bibinfo  {journal} {Physics Letters B}\ }\textbf {\bibinfo
  {volume} {761}},\ \bibinfo {pages} {125} (\bibinfo {year}
  {2016})}\BibitemShut {NoStop}%
\bibitem [{\citenamefont {Ramos}\ \emph {et~al.}(2018)\citenamefont {Ramos},
  \citenamefont {Caama{\~n}o}, \citenamefont {Farget}, \citenamefont
  {Rodr{\'i}guez-Tajes}, \citenamefont {Audouin}, \citenamefont {Benlliure},
  \citenamefont {Casarejos}, \citenamefont {Clement}, \citenamefont {Cortina},
  \citenamefont {Delaune}, \citenamefont {Derkx}, \citenamefont {Dijon},
  \citenamefont {Dor{\'e}}, \citenamefont {Fern{\'a}ndez-Dom{\'i}nguez},
  \citenamefont {de~France}, \citenamefont {Heinz}, \citenamefont {Jacquot},
  \citenamefont {Navin}, \citenamefont {Paradela}, \citenamefont {Rejmund},
  \citenamefont {Roger}, \citenamefont {Salsac},\ and\ \citenamefont
  {Schmitt}}]{ramPRC18}%
  \BibitemOpen
  \bibfield  {author} {\bibinfo {author} {\bibfnamefont {D.}~\bibnamefont
  {Ramos}}, \bibinfo {author} {\bibfnamefont {M.}~\bibnamefont {Caama{\~n}o}},
  \bibinfo {author} {\bibfnamefont {F.}~\bibnamefont {Farget}}, \bibinfo
  {author} {\bibfnamefont {C.}~\bibnamefont {Rodr{\'i}guez-Tajes}}, \bibinfo
  {author} {\bibfnamefont {L.}~\bibnamefont {Audouin}}, \bibinfo {author}
  {\bibfnamefont {J.}~\bibnamefont {Benlliure}}, \bibinfo {author}
  {\bibfnamefont {E.}~\bibnamefont {Casarejos}}, \bibinfo {author}
  {\bibfnamefont {E.}~\bibnamefont {Clement}}, \bibinfo {author} {\bibfnamefont
  {D.}~\bibnamefont {Cortina}}, \bibinfo {author} {\bibfnamefont
  {O.}~\bibnamefont {Delaune}}, \bibinfo {author} {\bibfnamefont
  {X.}~\bibnamefont {Derkx}}, \bibinfo {author} {\bibfnamefont
  {A.}~\bibnamefont {Dijon}}, \bibinfo {author} {\bibfnamefont
  {D.}~\bibnamefont {Dor{\'e}}}, \bibinfo {author} {\bibfnamefont
  {B.}~\bibnamefont {Fern{\'a}ndez-Dom{\'i}nguez}}, \bibinfo {author}
  {\bibfnamefont {G.}~\bibnamefont {de~France}}, \bibinfo {author}
  {\bibfnamefont {A.}~\bibnamefont {Heinz}}, \bibinfo {author} {\bibfnamefont
  {B.}~\bibnamefont {Jacquot}}, \bibinfo {author} {\bibfnamefont
  {A.}~\bibnamefont {Navin}}, \bibinfo {author} {\bibfnamefont
  {C.}~\bibnamefont {Paradela}}, \bibinfo {author} {\bibfnamefont
  {M.}~\bibnamefont {Rejmund}}, \bibinfo {author} {\bibfnamefont
  {T.}~\bibnamefont {Roger}}, \bibinfo {author} {\bibfnamefont {M.-D.}\
  \bibnamefont {Salsac}},\ and\ \bibinfo {author} {\bibfnamefont
  {C.}~\bibnamefont {Schmitt}},\ }\href
  {https://doi.org/10.1103/PhysRevC.97.054612} {\bibfield  {journal} {\bibinfo
  {journal} {Phys. Rev. C}\ }\textbf {\bibinfo {volume} {97}},\ \bibinfo
  {pages} {054612} (\bibinfo {year} {2018})}\BibitemShut {NoStop}%
\bibitem [{\citenamefont {Ward}\ \emph {et~al.}(2017)\citenamefont {Ward},
  \citenamefont {Carlsson}, \citenamefont {D{\o}ssing}, \citenamefont
  {M{\"{o}}ller}, \citenamefont {Randrup},\ and\ \citenamefont
  {{\AA}berg}}]{Ward2017}%
  \BibitemOpen
  \bibfield  {author} {\bibinfo {author} {\bibfnamefont {D.~E.}\ \bibnamefont
  {Ward}}, \bibinfo {author} {\bibfnamefont {B.~G.}\ \bibnamefont {Carlsson}},
  \bibinfo {author} {\bibfnamefont {T.}~\bibnamefont {D{\o}ssing}}, \bibinfo
  {author} {\bibfnamefont {P.}~\bibnamefont {M{\"{o}}ller}}, \bibinfo {author}
  {\bibfnamefont {J.}~\bibnamefont {Randrup}},\ and\ \bibinfo {author}
  {\bibfnamefont {S.}~\bibnamefont {{\AA}berg}},\ }\href
  {https://doi.org/10.1103/PhysRevC.95.024618} {\bibfield  {journal} {\bibinfo
  {journal} {Phys. Rev. C}\ }\textbf {\bibinfo {volume} {95}},\ \bibinfo
  {pages} {024618} (\bibinfo {year} {2017})}\BibitemShut {NoStop}%
\bibitem [{\citenamefont {Vermeulen}\ \emph {et~al.}(2020)\citenamefont
  {Vermeulen}, \citenamefont {Nishio}, \citenamefont {Hirose}, \citenamefont
  {Kean}, \citenamefont {Makii}, \citenamefont {Orlandi}, \citenamefont
  {Tsukada}, \citenamefont {Tsekhanovich}, \citenamefont {Andreyev},
  \citenamefont {Ishizaki}, \citenamefont {Okubayashi}, \citenamefont
  {Tanaka},\ and\ \citenamefont {Aritomo}}]{Nishio20}%
  \BibitemOpen
  \bibfield  {author} {\bibinfo {author} {\bibfnamefont {M.~J.}\ \bibnamefont
  {Vermeulen}}, \bibinfo {author} {\bibfnamefont {K.}~\bibnamefont {Nishio}},
  \bibinfo {author} {\bibfnamefont {K.}~\bibnamefont {Hirose}}, \bibinfo
  {author} {\bibfnamefont {K.~R.}\ \bibnamefont {Kean}}, \bibinfo {author}
  {\bibfnamefont {H.}~\bibnamefont {Makii}}, \bibinfo {author} {\bibfnamefont
  {R.}~\bibnamefont {Orlandi}}, \bibinfo {author} {\bibfnamefont
  {K.}~\bibnamefont {Tsukada}}, \bibinfo {author} {\bibfnamefont
  {I.}~\bibnamefont {Tsekhanovich}}, \bibinfo {author} {\bibfnamefont {A.~N.}\
  \bibnamefont {Andreyev}}, \bibinfo {author} {\bibfnamefont {S.}~\bibnamefont
  {Ishizaki}}, \bibinfo {author} {\bibfnamefont {M.}~\bibnamefont
  {Okubayashi}}, \bibinfo {author} {\bibfnamefont {S.}~\bibnamefont {Tanaka}},\
  and\ \bibinfo {author} {\bibfnamefont {Y.}~\bibnamefont {Aritomo}},\ }\href
  {https://doi.org/10.1103/PhysRevC.102.054610} {\bibfield  {journal} {\bibinfo
   {journal} {Phys. Rev. C}\ }\textbf {\bibinfo {volume} {102}},\ \bibinfo
  {pages} {054610} (\bibinfo {year} {2020})}\BibitemShut {NoStop}%
\bibitem [{\citenamefont {Schmidt}\ \emph {et~al.}(2000)\citenamefont
  {Schmidt}, \citenamefont {Steinh{\"{a}}user}, \citenamefont
  {B{\"{o}}ckstiegel}, \citenamefont {Grewe}, \citenamefont {Heinz},
  \citenamefont {Junghans}, \citenamefont {Benlliure}, \citenamefont {Clerc},
  \citenamefont {de~Jong}, \citenamefont {M{\"{u}}ller}, \citenamefont
  {Pf{\"{u}}tzner},\ and\ \citenamefont {Voss}}]{Sch00}%
  \BibitemOpen
  \bibfield  {author} {\bibinfo {author} {\bibfnamefont {K.-H.}\ \bibnamefont
  {Schmidt}}, \bibinfo {author} {\bibfnamefont {S.}~\bibnamefont
  {Steinh{\"{a}}user}}, \bibinfo {author} {\bibfnamefont {C.}~\bibnamefont
  {B{\"{o}}ckstiegel}}, \bibinfo {author} {\bibfnamefont {A.}~\bibnamefont
  {Grewe}}, \bibinfo {author} {\bibfnamefont {A.}~\bibnamefont {Heinz}},
  \bibinfo {author} {\bibfnamefont {A.}~\bibnamefont {Junghans}}, \bibinfo
  {author} {\bibfnamefont {J.}~\bibnamefont {Benlliure}}, \bibinfo {author}
  {\bibfnamefont {H.-G.}\ \bibnamefont {Clerc}}, \bibinfo {author}
  {\bibfnamefont {M.}~\bibnamefont {de~Jong}}, \bibinfo {author} {\bibfnamefont
  {J.}~\bibnamefont {M{\"{u}}ller}}, \bibinfo {author} {\bibfnamefont
  {M.}~\bibnamefont {Pf{\"{u}}tzner}},\ and\ \bibinfo {author} {\bibfnamefont
  {B.}~\bibnamefont {Voss}},\ }\href
  {https://doi.org/10.1016/S0375-9474(99)00384-X} {\bibfield  {journal}
  {\bibinfo  {journal} {Nucl. Phys. A}\ }\textbf {\bibinfo {volume} {665}},\
  \bibinfo {pages} {221} (\bibinfo {year} {2000})}\BibitemShut {NoStop}%
\bibitem [{\citenamefont {Caama{\~n}o}\ \emph {et~al.}(2013)\citenamefont
  {Caama{\~n}o}, \citenamefont {Delaune}, \citenamefont {Farget}, \citenamefont
  {Derkx}, \citenamefont {Schmidt}, \citenamefont {Audouin}, \citenamefont
  {Bacri}, \citenamefont {Barreau}, \citenamefont {Benlliure}, \citenamefont
  {Casarejos}, \citenamefont {Chbihi}, \citenamefont
  {Fern{\'a}ndez-Dom{\'i}nguez}, \citenamefont {Gaudefroy}, \citenamefont
  {Golabek}, \citenamefont {Jurado}, \citenamefont {Lemasson}, \citenamefont
  {Navin}, \citenamefont {Rejmund}, \citenamefont {Roger}, \citenamefont
  {Shrivastava},\ and\ \citenamefont {Schmitt}}]{caaPRC13}%
  \BibitemOpen
  \bibfield  {author} {\bibinfo {author} {\bibfnamefont {M.}~\bibnamefont
  {Caama{\~n}o}}, \bibinfo {author} {\bibfnamefont {O.}~\bibnamefont
  {Delaune}}, \bibinfo {author} {\bibfnamefont {F.}~\bibnamefont {Farget}},
  \bibinfo {author} {\bibfnamefont {X.}~\bibnamefont {Derkx}}, \bibinfo
  {author} {\bibfnamefont {K.-H.}\ \bibnamefont {Schmidt}}, \bibinfo {author}
  {\bibfnamefont {L.}~\bibnamefont {Audouin}}, \bibinfo {author} {\bibfnamefont
  {C.-O.}\ \bibnamefont {Bacri}}, \bibinfo {author} {\bibfnamefont
  {G.}~\bibnamefont {Barreau}}, \bibinfo {author} {\bibfnamefont
  {J.}~\bibnamefont {Benlliure}}, \bibinfo {author} {\bibfnamefont
  {E.}~\bibnamefont {Casarejos}}, \bibinfo {author} {\bibfnamefont
  {A.}~\bibnamefont {Chbihi}}, \bibinfo {author} {\bibfnamefont
  {B.}~\bibnamefont {Fern{\'a}ndez-Dom{\'i}nguez}}, \bibinfo {author}
  {\bibfnamefont {L.}~\bibnamefont {Gaudefroy}}, \bibinfo {author}
  {\bibfnamefont {C.}~\bibnamefont {Golabek}}, \bibinfo {author} {\bibfnamefont
  {B.}~\bibnamefont {Jurado}}, \bibinfo {author} {\bibfnamefont
  {A.}~\bibnamefont {Lemasson}}, \bibinfo {author} {\bibfnamefont
  {A.}~\bibnamefont {Navin}}, \bibinfo {author} {\bibfnamefont
  {M.}~\bibnamefont {Rejmund}}, \bibinfo {author} {\bibfnamefont
  {T.}~\bibnamefont {Roger}}, \bibinfo {author} {\bibfnamefont
  {A.}~\bibnamefont {Shrivastava}},\ and\ \bibinfo {author} {\bibfnamefont
  {C.}~\bibnamefont {Schmitt}},\ }\href
  {https://doi.org/10.1103/PhysRevC.88.024605} {\bibfield  {journal} {\bibinfo
  {journal} {Phys. Rev. C}\ }\textbf {\bibinfo {volume} {88}},\ \bibinfo
  {pages} {024605} (\bibinfo {year} {2013})}\BibitemShut {NoStop}%
\bibitem [{\citenamefont {Pellereau}\ \emph {et~al.}(2017)\citenamefont
  {Pellereau}, \citenamefont {Ta\"{i}eb}, \citenamefont {Chatillon},
  \citenamefont {Alvarez-Pol}, \citenamefont {Audouin}, \citenamefont {Ayyad},
  \citenamefont {B\'elier}, \citenamefont {Benlliure}, \citenamefont {Boutoux},
  \citenamefont {Caama\~no}, \citenamefont {Casarejos}, \citenamefont
  {Cortina-Gil}, \citenamefont {Ebran}, \citenamefont {Farget}, \citenamefont
  {Fern\'andez-Dom\'{i}nguez}, \citenamefont {Gorbinet}, \citenamefont
  {Grente}, \citenamefont {Heinz}, \citenamefont {Johansson}, \citenamefont
  {Jurado}, \citenamefont {Keli\ifmmode \acute{c}\else~\'{c}\fi{} Heil},
  \citenamefont {Kurz}, \citenamefont {Laurent}, \citenamefont {Martin},
  \citenamefont {Nociforo}, \citenamefont {Paradela}, \citenamefont {Pietri},
  \citenamefont {Rodr\'{i}guez-S\'anchez}, \citenamefont {Schmidt},
  \citenamefont {Simon}, \citenamefont {Tassan-Got}, \citenamefont {Vargas},
  \citenamefont {Voss},\ and\ \citenamefont {Weick}}]{pell17}%
  \BibitemOpen
  \bibfield  {author} {\bibinfo {author} {\bibfnamefont {E.}~\bibnamefont
  {Pellereau}}, \bibinfo {author} {\bibfnamefont {J.}~\bibnamefont
  {Ta\"{i}eb}}, \bibinfo {author} {\bibfnamefont {A.}~\bibnamefont
  {Chatillon}}, \bibinfo {author} {\bibfnamefont {H.}~\bibnamefont
  {Alvarez-Pol}}, \bibinfo {author} {\bibfnamefont {L.}~\bibnamefont
  {Audouin}}, \bibinfo {author} {\bibfnamefont {Y.}~\bibnamefont {Ayyad}},
  \bibinfo {author} {\bibfnamefont {G.}~\bibnamefont {B\'elier}}, \bibinfo
  {author} {\bibfnamefont {J.}~\bibnamefont {Benlliure}}, \bibinfo {author}
  {\bibfnamefont {G.}~\bibnamefont {Boutoux}}, \bibinfo {author} {\bibfnamefont
  {M.}~\bibnamefont {Caama\~no}}, \bibinfo {author} {\bibfnamefont
  {E.}~\bibnamefont {Casarejos}}, \bibinfo {author} {\bibfnamefont
  {D.}~\bibnamefont {Cortina-Gil}}, \bibinfo {author} {\bibfnamefont
  {A.}~\bibnamefont {Ebran}}, \bibinfo {author} {\bibfnamefont
  {F.}~\bibnamefont {Farget}}, \bibinfo {author} {\bibfnamefont
  {B.}~\bibnamefont {Fern\'andez-Dom\'{i}nguez}}, \bibinfo {author}
  {\bibfnamefont {T.}~\bibnamefont {Gorbinet}}, \bibinfo {author}
  {\bibfnamefont {L.}~\bibnamefont {Grente}}, \bibinfo {author} {\bibfnamefont
  {A.}~\bibnamefont {Heinz}}, \bibinfo {author} {\bibfnamefont
  {H.}~\bibnamefont {Johansson}}, \bibinfo {author} {\bibfnamefont
  {B.}~\bibnamefont {Jurado}}, \bibinfo {author} {\bibfnamefont
  {A.}~\bibnamefont {Keli\ifmmode \acute{c}\else~\'{c}\fi{} Heil}}, \bibinfo
  {author} {\bibfnamefont {N.}~\bibnamefont {Kurz}}, \bibinfo {author}
  {\bibfnamefont {B.}~\bibnamefont {Laurent}}, \bibinfo {author} {\bibfnamefont
  {J.-F.}\ \bibnamefont {Martin}}, \bibinfo {author} {\bibfnamefont
  {C.}~\bibnamefont {Nociforo}}, \bibinfo {author} {\bibfnamefont
  {C.}~\bibnamefont {Paradela}}, \bibinfo {author} {\bibfnamefont
  {S.}~\bibnamefont {Pietri}}, \bibinfo {author} {\bibfnamefont {J.~L.}\
  \bibnamefont {Rodr\'{i}guez-S\'anchez}}, \bibinfo {author} {\bibfnamefont
  {K.-H.}\ \bibnamefont {Schmidt}}, \bibinfo {author} {\bibfnamefont
  {H.}~\bibnamefont {Simon}}, \bibinfo {author} {\bibfnamefont
  {L.}~\bibnamefont {Tassan-Got}}, \bibinfo {author} {\bibfnamefont
  {J.}~\bibnamefont {Vargas}}, \bibinfo {author} {\bibfnamefont
  {B.}~\bibnamefont {Voss}},\ and\ \bibinfo {author} {\bibfnamefont
  {H.}~\bibnamefont {Weick}},\ }\href
  {https://doi.org/10.1103/PhysRevC.95.054603} {\bibfield  {journal} {\bibinfo
  {journal} {Phys. Rev. C}\ }\textbf {\bibinfo {volume} {95}},\ \bibinfo
  {pages} {054603} (\bibinfo {year} {2017})}\BibitemShut {NoStop}%
\bibitem [{\citenamefont {Ramos}\ \emph {et~al.}(2019)\citenamefont {Ramos},
  \citenamefont {Caama\~no}, \citenamefont {Lemasson}, \citenamefont {Rejmund},
  \citenamefont {Audouin}, \citenamefont {\'Alvarez-Pol}, \citenamefont
  {Frankland}, \citenamefont {Fern\'andez-Dom{\'i}nguez}, \citenamefont
  {Galiana-Bald\'o}, \citenamefont {Piot}, \citenamefont {Ackermann},
  \citenamefont {Biswas}, \citenamefont {Clement}, \citenamefont {Durand},
  \citenamefont {Farget}, \citenamefont {Fregeau}, \citenamefont {Galaviz},
  \citenamefont {Heinz}, \citenamefont {Henriques}, \citenamefont {Jacquot},
  \citenamefont {Jurado}, \citenamefont {Kim}, \citenamefont {Morfouace},
  \citenamefont {Ralet}, \citenamefont {Roger}, \citenamefont {Schmitt},
  \citenamefont {Teubig},\ and\ \citenamefont {Tsekhanovich}}]{ramPRL}%
  \BibitemOpen
  \bibfield  {author} {\bibinfo {author} {\bibfnamefont {D.}~\bibnamefont
  {Ramos}}, \bibinfo {author} {\bibfnamefont {M.}~\bibnamefont {Caama\~no}},
  \bibinfo {author} {\bibfnamefont {A.}~\bibnamefont {Lemasson}}, \bibinfo
  {author} {\bibfnamefont {M.}~\bibnamefont {Rejmund}}, \bibinfo {author}
  {\bibfnamefont {L.}~\bibnamefont {Audouin}}, \bibinfo {author} {\bibfnamefont
  {H.}~\bibnamefont {\'Alvarez-Pol}}, \bibinfo {author} {\bibfnamefont {J.~D.}\
  \bibnamefont {Frankland}}, \bibinfo {author} {\bibfnamefont {B.}~\bibnamefont
  {Fern\'andez-Dom{\'i}nguez}}, \bibinfo {author} {\bibfnamefont
  {E.}~\bibnamefont {Galiana-Bald\'o}}, \bibinfo {author} {\bibfnamefont
  {J.}~\bibnamefont {Piot}}, \bibinfo {author} {\bibfnamefont {D.}~\bibnamefont
  {Ackermann}}, \bibinfo {author} {\bibfnamefont {S.}~\bibnamefont {Biswas}},
  \bibinfo {author} {\bibfnamefont {E.}~\bibnamefont {Clement}}, \bibinfo
  {author} {\bibfnamefont {D.}~\bibnamefont {Durand}}, \bibinfo {author}
  {\bibfnamefont {F.}~\bibnamefont {Farget}}, \bibinfo {author} {\bibfnamefont
  {M.~O.}\ \bibnamefont {Fregeau}}, \bibinfo {author} {\bibfnamefont
  {D.}~\bibnamefont {Galaviz}}, \bibinfo {author} {\bibfnamefont
  {A.}~\bibnamefont {Heinz}}, \bibinfo {author} {\bibfnamefont {A.~I.}\
  \bibnamefont {Henriques}}, \bibinfo {author} {\bibfnamefont {B.}~\bibnamefont
  {Jacquot}}, \bibinfo {author} {\bibfnamefont {B.}~\bibnamefont {Jurado}},
  \bibinfo {author} {\bibfnamefont {Y.~H.}\ \bibnamefont {Kim}}, \bibinfo
  {author} {\bibfnamefont {P.}~\bibnamefont {Morfouace}}, \bibinfo {author}
  {\bibfnamefont {D.}~\bibnamefont {Ralet}}, \bibinfo {author} {\bibfnamefont
  {T.}~\bibnamefont {Roger}}, \bibinfo {author} {\bibfnamefont
  {C.}~\bibnamefont {Schmitt}}, \bibinfo {author} {\bibfnamefont
  {P.}~\bibnamefont {Teubig}},\ and\ \bibinfo {author} {\bibfnamefont
  {I.}~\bibnamefont {Tsekhanovich}},\ }\href
  {https://doi.org/10.1103/PhysRevLett.123.092503} {\bibfield  {journal}
  {\bibinfo  {journal} {Phys. Rev. Lett.}\ }\textbf {\bibinfo {volume} {123}},\
  \bibinfo {pages} {092503} (\bibinfo {year} {2019})}\BibitemShut {NoStop}%
\bibitem [{\citenamefont {Chatillon}\ \emph {et~al.}(2020)\citenamefont
  {Chatillon}, \citenamefont {Ta\"{i}eb}, \citenamefont {Alvarez-Pol},
  \citenamefont {Audouin}, \citenamefont {Ayyad}, \citenamefont {B\'elier},
  \citenamefont {Benlliure}, \citenamefont {Boutoux}, \citenamefont
  {Caama\~no}, \citenamefont {Casarejos}, \citenamefont {Cortina-Gil},
  \citenamefont {Ebran}, \citenamefont {Farget}, \citenamefont
  {Fern\'andez-Dom\'{i}nguez}, \citenamefont {Gorbinet}, \citenamefont
  {Grente}, \citenamefont {Heinz}, \citenamefont {Johansson}, \citenamefont
  {Jurado}, \citenamefont {Keli\ifmmode \acute{c}\else~\'{c}\fi{} Heil},
  \citenamefont {Kurz}, \citenamefont {Laurent}, \citenamefont {Martin},
  \citenamefont {Nociforo}, \citenamefont {Paradela}, \citenamefont
  {Pellereau}, \citenamefont {Pietri}, \citenamefont {Prochazka}, \citenamefont
  {Rodr\'{i}guez-S\'anchez}, \citenamefont {Rossi}, \citenamefont {Simon},
  \citenamefont {Tassan-Got}, \citenamefont {Vargas}, \citenamefont {Voss},\
  and\ \citenamefont {Weick}}]{chat20}%
  \BibitemOpen
  \bibfield  {author} {\bibinfo {author} {\bibfnamefont {A.}~\bibnamefont
  {Chatillon}}, \bibinfo {author} {\bibfnamefont {J.}~\bibnamefont
  {Ta\"{i}eb}}, \bibinfo {author} {\bibfnamefont {H.}~\bibnamefont
  {Alvarez-Pol}}, \bibinfo {author} {\bibfnamefont {L.}~\bibnamefont
  {Audouin}}, \bibinfo {author} {\bibfnamefont {Y.}~\bibnamefont {Ayyad}},
  \bibinfo {author} {\bibfnamefont {G.}~\bibnamefont {B\'elier}}, \bibinfo
  {author} {\bibfnamefont {J.}~\bibnamefont {Benlliure}}, \bibinfo {author}
  {\bibfnamefont {G.}~\bibnamefont {Boutoux}}, \bibinfo {author} {\bibfnamefont
  {M.}~\bibnamefont {Caama\~no}}, \bibinfo {author} {\bibfnamefont
  {E.}~\bibnamefont {Casarejos}}, \bibinfo {author} {\bibfnamefont
  {D.}~\bibnamefont {Cortina-Gil}}, \bibinfo {author} {\bibfnamefont
  {A.}~\bibnamefont {Ebran}}, \bibinfo {author} {\bibfnamefont
  {F.}~\bibnamefont {Farget}}, \bibinfo {author} {\bibfnamefont
  {B.}~\bibnamefont {Fern\'andez-Dom\'{i}nguez}}, \bibinfo {author}
  {\bibfnamefont {T.}~\bibnamefont {Gorbinet}}, \bibinfo {author}
  {\bibfnamefont {L.}~\bibnamefont {Grente}}, \bibinfo {author} {\bibfnamefont
  {A.}~\bibnamefont {Heinz}}, \bibinfo {author} {\bibfnamefont {H.~T.}\
  \bibnamefont {Johansson}}, \bibinfo {author} {\bibfnamefont {B.}~\bibnamefont
  {Jurado}}, \bibinfo {author} {\bibfnamefont {A.}~\bibnamefont {Keli\ifmmode
  \acute{c}\else~\'{c}\fi{} Heil}}, \bibinfo {author} {\bibfnamefont
  {N.}~\bibnamefont {Kurz}}, \bibinfo {author} {\bibfnamefont {B.}~\bibnamefont
  {Laurent}}, \bibinfo {author} {\bibfnamefont {J.-F.}\ \bibnamefont {Martin}},
  \bibinfo {author} {\bibfnamefont {C.}~\bibnamefont {Nociforo}}, \bibinfo
  {author} {\bibfnamefont {C.}~\bibnamefont {Paradela}}, \bibinfo {author}
  {\bibfnamefont {E.}~\bibnamefont {Pellereau}}, \bibinfo {author}
  {\bibfnamefont {S.}~\bibnamefont {Pietri}}, \bibinfo {author} {\bibfnamefont
  {A.}~\bibnamefont {Prochazka}}, \bibinfo {author} {\bibfnamefont {J.~L.}\
  \bibnamefont {Rodr\'{i}guez-S\'anchez}}, \bibinfo {author} {\bibfnamefont
  {D.}~\bibnamefont {Rossi}}, \bibinfo {author} {\bibfnamefont
  {H.}~\bibnamefont {Simon}}, \bibinfo {author} {\bibfnamefont
  {L.}~\bibnamefont {Tassan-Got}}, \bibinfo {author} {\bibfnamefont
  {J.}~\bibnamefont {Vargas}}, \bibinfo {author} {\bibfnamefont
  {B.}~\bibnamefont {Voss}},\ and\ \bibinfo {author} {\bibfnamefont
  {H.}~\bibnamefont {Weick}},\ }\href
  {https://doi.org/10.1103/PhysRevLett.124.202502} {\bibfield  {journal}
  {\bibinfo  {journal} {Phys. Rev. Lett.}\ }\textbf {\bibinfo {volume} {124}},\
  \bibinfo {pages} {202502} (\bibinfo {year} {2020})}\BibitemShut {NoStop}%
\bibitem [{\citenamefont {Rodr{\'{i}}guez-Tajes}\ \emph
  {et~al.}(2014)\citenamefont {Rodr{\'{i}}guez-Tajes}, \citenamefont {Farget},
  \citenamefont {Derkx}, \citenamefont {Caama{\~{n}}o}, \citenamefont
  {Delaune}, \citenamefont {Schmidt}, \citenamefont {Cl{\'{e}}ment},
  \citenamefont {Dijon}, \citenamefont {Heinz}, \citenamefont {Roger},
  \citenamefont {Audouin}, \citenamefont {Benlliure}, \citenamefont
  {Casarejos}, \citenamefont {Cortina}, \citenamefont {Dor{\'{e}}},
  \citenamefont {Fern{\'{a}}ndez-Dom{\'{i}}nguez}, \citenamefont {Jacquot},
  \citenamefont {Jurado}, \citenamefont {Navin}, \citenamefont {Paradela},
  \citenamefont {Ramos}, \citenamefont {Romain}, \citenamefont {Salsac},\ and\
  \citenamefont {Schmitt}}]{rodPRC14}%
  \BibitemOpen
  \bibfield  {author} {\bibinfo {author} {\bibfnamefont {C.}~\bibnamefont
  {Rodr{\'{i}}guez-Tajes}}, \bibinfo {author} {\bibfnamefont {F.}~\bibnamefont
  {Farget}}, \bibinfo {author} {\bibfnamefont {X.}~\bibnamefont {Derkx}},
  \bibinfo {author} {\bibfnamefont {M.}~\bibnamefont {Caama{\~{n}}o}}, \bibinfo
  {author} {\bibfnamefont {O.}~\bibnamefont {Delaune}}, \bibinfo {author}
  {\bibfnamefont {K.-H.}\ \bibnamefont {Schmidt}}, \bibinfo {author}
  {\bibfnamefont {E.}~\bibnamefont {Cl{\'{e}}ment}}, \bibinfo {author}
  {\bibfnamefont {A.}~\bibnamefont {Dijon}}, \bibinfo {author} {\bibfnamefont
  {A.}~\bibnamefont {Heinz}}, \bibinfo {author} {\bibfnamefont
  {T.}~\bibnamefont {Roger}}, \bibinfo {author} {\bibfnamefont
  {L.}~\bibnamefont {Audouin}}, \bibinfo {author} {\bibfnamefont
  {J.}~\bibnamefont {Benlliure}}, \bibinfo {author} {\bibfnamefont
  {E.}~\bibnamefont {Casarejos}}, \bibinfo {author} {\bibfnamefont
  {D.}~\bibnamefont {Cortina}}, \bibinfo {author} {\bibfnamefont
  {D.}~\bibnamefont {Dor{\'{e}}}}, \bibinfo {author} {\bibfnamefont
  {B.}~\bibnamefont {Fern{\'{a}}ndez-Dom{\'{i}}nguez}}, \bibinfo {author}
  {\bibfnamefont {B.}~\bibnamefont {Jacquot}}, \bibinfo {author} {\bibfnamefont
  {B.}~\bibnamefont {Jurado}}, \bibinfo {author} {\bibfnamefont
  {A.}~\bibnamefont {Navin}}, \bibinfo {author} {\bibfnamefont
  {C.}~\bibnamefont {Paradela}}, \bibinfo {author} {\bibfnamefont
  {D.}~\bibnamefont {Ramos}}, \bibinfo {author} {\bibfnamefont
  {P.}~\bibnamefont {Romain}}, \bibinfo {author} {\bibfnamefont {M.~D.}\
  \bibnamefont {Salsac}},\ and\ \bibinfo {author} {\bibfnamefont
  {C.}~\bibnamefont {Schmitt}},\ }\href
  {https://doi.org/10.1103/PhysRevC.89.024614} {\bibfield  {journal} {\bibinfo
  {journal} {Phys. Rev. C}\ }\textbf {\bibinfo {volume} {89}},\ \bibinfo
  {pages} {024614} (\bibinfo {year} {2014})}\BibitemShut {NoStop}%
\bibitem [{\citenamefont {Rejmund}\ \emph {et~al.}(2011)\citenamefont
  {Rejmund}, \citenamefont {Lecornu}, \citenamefont {Navin}, \citenamefont
  {Schmitt}, \citenamefont {Damoy}, \citenamefont {Delaune}, \citenamefont
  {Enguerrand}, \citenamefont {Fremont}, \citenamefont {Gangnant},
  \citenamefont {Gaudefroy}, \citenamefont {Jacquot}, \citenamefont {Pancin},
  \citenamefont {Pullanhiotan},\ and\ \citenamefont {Spitaels}}]{rejNIM}%
  \BibitemOpen
  \bibfield  {author} {\bibinfo {author} {\bibfnamefont {M.}~\bibnamefont
  {Rejmund}}, \bibinfo {author} {\bibfnamefont {B.}~\bibnamefont {Lecornu}},
  \bibinfo {author} {\bibfnamefont {A.}~\bibnamefont {Navin}}, \bibinfo
  {author} {\bibfnamefont {C.}~\bibnamefont {Schmitt}}, \bibinfo {author}
  {\bibfnamefont {S.}~\bibnamefont {Damoy}}, \bibinfo {author} {\bibfnamefont
  {O.}~\bibnamefont {Delaune}}, \bibinfo {author} {\bibfnamefont {J.~M.}\
  \bibnamefont {Enguerrand}}, \bibinfo {author} {\bibfnamefont
  {G.}~\bibnamefont {Fremont}}, \bibinfo {author} {\bibfnamefont
  {P.}~\bibnamefont {Gangnant}}, \bibinfo {author} {\bibfnamefont
  {L.}~\bibnamefont {Gaudefroy}}, \bibinfo {author} {\bibfnamefont
  {B.}~\bibnamefont {Jacquot}}, \bibinfo {author} {\bibfnamefont
  {J.}~\bibnamefont {Pancin}}, \bibinfo {author} {\bibfnamefont
  {S.}~\bibnamefont {Pullanhiotan}},\ and\ \bibinfo {author} {\bibfnamefont
  {C.}~\bibnamefont {Spitaels}},\ }\href
  {https://doi.org/10.1016/j.nima.2011.05.007} {\bibfield  {journal} {\bibinfo
  {journal} {Nucl. Instr. Meth. A}\ }\textbf {\bibinfo {volume} {646}},\
  \bibinfo {pages} {184} (\bibinfo {year} {2011})}\BibitemShut {NoStop}%
\bibitem [{\citenamefont {Lemasson}\ and\ \citenamefont
  {Rejmund}(2023)}]{LemNIM}%
  \BibitemOpen
  \bibfield  {author} {\bibinfo {author} {\bibfnamefont {A.}~\bibnamefont
  {Lemasson}}\ and\ \bibinfo {author} {\bibfnamefont {M.}~\bibnamefont
  {Rejmund}},\ }\href
  {https://doi.org/https://doi.org/10.1016/j.nima.2023.168407} {\bibfield
  {journal} {\bibinfo  {journal} {Nuclear Instruments and Methods in Physics
  Research Section A: Accelerators, Spectrometers, Detectors and Associated
  Equipment}\ }\textbf {\bibinfo {volume} {1054}},\ \bibinfo {pages} {168407}
  (\bibinfo {year} {2023})}\BibitemShut {NoStop}%
\bibitem [{\citenamefont {Gindler}\ \emph {et~al.}(1983)\citenamefont
  {Gindler}, \citenamefont {Glendenin}, \citenamefont {Henderson},\ and\
  \citenamefont {Meadows}}]{GinPRC}%
  \BibitemOpen
  \bibfield  {author} {\bibinfo {author} {\bibfnamefont {J.~E.}\ \bibnamefont
  {Gindler}}, \bibinfo {author} {\bibfnamefont {L.~E.}\ \bibnamefont
  {Glendenin}}, \bibinfo {author} {\bibfnamefont {D.~J.}\ \bibnamefont
  {Henderson}},\ and\ \bibinfo {author} {\bibfnamefont {J.~W.}\ \bibnamefont
  {Meadows}},\ }\href {https://doi.org/10.1103/PhysRevC.27.2058} {\bibfield
  {journal} {\bibinfo  {journal} {Phys. Rev. C}\ }\textbf {\bibinfo {volume}
  {27}},\ \bibinfo {pages} {2058} (\bibinfo {year} {1983})}\BibitemShut
  {NoStop}%
\bibitem [{\citenamefont {Schmitt}\ \emph {et~al.}(1984)\citenamefont
  {Schmitt}, \citenamefont {Guessous}, \citenamefont {Bocquet}, \citenamefont
  {Clerc}, \citenamefont {Brissot}, \citenamefont {Engelhardt}, \citenamefont
  {Faust}, \citenamefont {Gönnenwein}, \citenamefont {Mutterer}, \citenamefont
  {Nifenecker}, \citenamefont {Pannicke}, \citenamefont {Ristori},\ and\
  \citenamefont {Theobald}}]{schnpa}%
  \BibitemOpen
  \bibfield  {author} {\bibinfo {author} {\bibfnamefont {C.}~\bibnamefont
  {Schmitt}}, \bibinfo {author} {\bibfnamefont {A.}~\bibnamefont {Guessous}},
  \bibinfo {author} {\bibfnamefont {J.}~\bibnamefont {Bocquet}}, \bibinfo
  {author} {\bibfnamefont {H.-G.}\ \bibnamefont {Clerc}}, \bibinfo {author}
  {\bibfnamefont {R.}~\bibnamefont {Brissot}}, \bibinfo {author} {\bibfnamefont
  {D.}~\bibnamefont {Engelhardt}}, \bibinfo {author} {\bibfnamefont
  {H.}~\bibnamefont {Faust}}, \bibinfo {author} {\bibfnamefont
  {F.}~\bibnamefont {Gönnenwein}}, \bibinfo {author} {\bibfnamefont
  {M.}~\bibnamefont {Mutterer}}, \bibinfo {author} {\bibfnamefont
  {H.}~\bibnamefont {Nifenecker}}, \bibinfo {author} {\bibfnamefont
  {J.}~\bibnamefont {Pannicke}}, \bibinfo {author} {\bibfnamefont
  {C.}~\bibnamefont {Ristori}},\ and\ \bibinfo {author} {\bibfnamefont
  {J.}~\bibnamefont {Theobald}},\ }\href
  {https://doi.org/https://doi.org/10.1016/0375-9474(84)90191-X} {\bibfield
  {journal} {\bibinfo  {journal} {Nuclear Physics A}\ }\textbf {\bibinfo
  {volume} {430}},\ \bibinfo {pages} {21} (\bibinfo {year} {1984})}\BibitemShut
  {NoStop}%
\bibitem [{\citenamefont {Bail}\ \emph {et~al.}(2011)\citenamefont {Bail},
  \citenamefont {Serot}, \citenamefont {Mathieu}, \citenamefont {Litaize},
  \citenamefont {Materna}, \citenamefont {K\"oster}, \citenamefont {Faust},
  \citenamefont {Letourneau},\ and\ \citenamefont {Panebianco}}]{BaiPRC}%
  \BibitemOpen
  \bibfield  {author} {\bibinfo {author} {\bibfnamefont {A.}~\bibnamefont
  {Bail}}, \bibinfo {author} {\bibfnamefont {O.}~\bibnamefont {Serot}},
  \bibinfo {author} {\bibfnamefont {L.}~\bibnamefont {Mathieu}}, \bibinfo
  {author} {\bibfnamefont {O.}~\bibnamefont {Litaize}}, \bibinfo {author}
  {\bibfnamefont {T.}~\bibnamefont {Materna}}, \bibinfo {author} {\bibfnamefont
  {U.}~\bibnamefont {K\"oster}}, \bibinfo {author} {\bibfnamefont
  {H.}~\bibnamefont {Faust}}, \bibinfo {author} {\bibfnamefont
  {A.}~\bibnamefont {Letourneau}},\ and\ \bibinfo {author} {\bibfnamefont
  {S.}~\bibnamefont {Panebianco}},\ }\href
  {https://doi.org/10.1103/PhysRevC.84.034605} {\bibfield  {journal} {\bibinfo
  {journal} {Phys. Rev. C}\ }\textbf {\bibinfo {volume} {84}},\ \bibinfo
  {pages} {034605} (\bibinfo {year} {2011})}\BibitemShut {NoStop}%
\bibitem [{\citenamefont {J.Frehaut}\ \emph {et~al.}(1973)\citenamefont
  {J.Frehaut}, \citenamefont {G.Mosinski}, \citenamefont {R.Bois},\ and\
  \citenamefont {M.Soleilhac}}]{Fre80}%
  \BibitemOpen
  \bibfield  {author} {\bibinfo {author} {\bibnamefont {J.Frehaut}}, \bibinfo
  {author} {\bibnamefont {G.Mosinski}}, \bibinfo {author} {\bibnamefont
  {R.Bois}},\ and\ \bibinfo {author} {\bibnamefont {M.Soleilhac}},\ }\href
  {https://www-nds.iaea.org/exfor/servlet/X4sShowPubl?File=C,73KIEV,3,153,1973}
  {\bibinfo {title} {Measurement of the average prompt neutron number nu-p
  emitted during fission of 240pu and 241pu induced by neutrons with energy
  between 1.5 and 15 mev}} (\bibinfo {year} {1973}),\ \bibinfo {note} {eANDC
  Topical Conference 154, 67. EXOR entry n. 20488}\BibitemShut {NoStop}%
\bibitem [{\citenamefont {Marini}\ \emph {et~al.}(2022)\citenamefont {Marini},
  \citenamefont {Taieb}, \citenamefont {Neudecker}, \citenamefont {Bélier},
  \citenamefont {Chatillon}, \citenamefont {Etasse}, \citenamefont {Laurent},
  \citenamefont {Morfouace}, \citenamefont {Morillon}, \citenamefont {Devlin},
  \citenamefont {Gomez}, \citenamefont {Haight}, \citenamefont {Kelly},\ and\
  \citenamefont {O'Donnell}}]{Mar22}%
  \BibitemOpen
  \bibfield  {author} {\bibinfo {author} {\bibfnamefont {P.}~\bibnamefont
  {Marini}}, \bibinfo {author} {\bibfnamefont {J.}~\bibnamefont {Taieb}},
  \bibinfo {author} {\bibfnamefont {D.}~\bibnamefont {Neudecker}}, \bibinfo
  {author} {\bibfnamefont {G.}~\bibnamefont {Bélier}}, \bibinfo {author}
  {\bibfnamefont {A.}~\bibnamefont {Chatillon}}, \bibinfo {author}
  {\bibfnamefont {D.}~\bibnamefont {Etasse}}, \bibinfo {author} {\bibfnamefont
  {B.}~\bibnamefont {Laurent}}, \bibinfo {author} {\bibfnamefont
  {P.}~\bibnamefont {Morfouace}}, \bibinfo {author} {\bibfnamefont
  {B.}~\bibnamefont {Morillon}}, \bibinfo {author} {\bibfnamefont
  {M.}~\bibnamefont {Devlin}}, \bibinfo {author} {\bibfnamefont
  {J.}~\bibnamefont {Gomez}}, \bibinfo {author} {\bibfnamefont
  {R.}~\bibnamefont {Haight}}, \bibinfo {author} {\bibfnamefont
  {K.}~\bibnamefont {Kelly}},\ and\ \bibinfo {author} {\bibfnamefont
  {J.}~\bibnamefont {O'Donnell}},\ }\href
  {https://doi.org/https://doi.org/10.1016/j.physletb.2022.137513} {\bibfield
  {journal} {\bibinfo  {journal} {Physics Letters B}\ }\textbf {\bibinfo
  {volume} {835}},\ \bibinfo {pages} {137513} (\bibinfo {year}
  {2022})}\BibitemShut {NoStop}%
\bibitem [{\citenamefont {Caldwell}\ \emph {et~al.}(1980)\citenamefont
  {Caldwell}, \citenamefont {Dowdy}, \citenamefont {Berman}, \citenamefont
  {Alvarez},\ and\ \citenamefont {Meyer}}]{Caldwell80}%
  \BibitemOpen
  \bibfield  {author} {\bibinfo {author} {\bibfnamefont {J.~T.}\ \bibnamefont
  {Caldwell}}, \bibinfo {author} {\bibfnamefont {E.~J.}\ \bibnamefont {Dowdy}},
  \bibinfo {author} {\bibfnamefont {B.~L.}\ \bibnamefont {Berman}}, \bibinfo
  {author} {\bibfnamefont {R.~A.}\ \bibnamefont {Alvarez}},\ and\ \bibinfo
  {author} {\bibfnamefont {P.}~\bibnamefont {Meyer}},\ }\href
  {https://doi.org/10.1103/PhysRevC.21.1215} {\bibfield  {journal} {\bibinfo
  {journal} {Phys. Rev. C}\ }\textbf {\bibinfo {volume} {21}},\ \bibinfo
  {pages} {1215} (\bibinfo {year} {1980})}\BibitemShut {NoStop}%
\bibitem [{\citenamefont {Naik}\ \emph {et~al.}(2015)\citenamefont {Naik},
  \citenamefont {Kim}, \citenamefont {Schwengner}, \citenamefont {Kim},
  \citenamefont {John}, \citenamefont {Massarczyk}, \citenamefont {Junghans},
  \citenamefont {Wagner},\ and\ \citenamefont {Goswami}}]{Naik15}%
  \BibitemOpen
  \bibfield  {author} {\bibinfo {author} {\bibfnamefont {H.}~\bibnamefont
  {Naik}}, \bibinfo {author} {\bibfnamefont {G.}~\bibnamefont {Kim}}, \bibinfo
  {author} {\bibfnamefont {R.}~\bibnamefont {Schwengner}}, \bibinfo {author}
  {\bibfnamefont {K.}~\bibnamefont {Kim}}, \bibinfo {author} {\bibfnamefont
  {R.}~\bibnamefont {John}}, \bibinfo {author} {\bibfnamefont {R.}~\bibnamefont
  {Massarczyk}}, \bibinfo {author} {\bibfnamefont {A.}~\bibnamefont
  {Junghans}}, \bibinfo {author} {\bibfnamefont {A.}~\bibnamefont {Wagner}},\
  and\ \bibinfo {author} {\bibfnamefont {A.}~\bibnamefont {Goswami}},\ }\href
  {https://doi.org/10.1140/epja/i2015-15150-8} {\bibfield  {journal} {\bibinfo
  {journal} {The European Physical Journal A}\ }\textbf {\bibinfo {volume}
  {51}},\ \bibinfo {pages} {150} (\bibinfo {year} {2015})}\BibitemShut
  {NoStop}%
\bibitem [{\citenamefont {Filipescu}\ \emph {et~al.}(2024)\citenamefont
  {Filipescu}, \citenamefont {Gheorghe}, \citenamefont {Goriely}, \citenamefont
  {Tudora}, \citenamefont {Nishio}, \citenamefont {Ohtsuki}, \citenamefont
  {Wang}, \citenamefont {Fan}, \citenamefont {Stopani}, \citenamefont {Suzaki},
  \citenamefont {Hirose}, \citenamefont {Inagaki}, \citenamefont {Lui},
  \citenamefont {Ari-izumi}, \citenamefont {Miyamoto}, \citenamefont {Otsuka},\
  and\ \citenamefont {Utsunomiya}}]{Filipescu24}%
  \BibitemOpen
  \bibfield  {author} {\bibinfo {author} {\bibfnamefont {D.}~\bibnamefont
  {Filipescu}}, \bibinfo {author} {\bibfnamefont {I.}~\bibnamefont {Gheorghe}},
  \bibinfo {author} {\bibfnamefont {S.}~\bibnamefont {Goriely}}, \bibinfo
  {author} {\bibfnamefont {A.}~\bibnamefont {Tudora}}, \bibinfo {author}
  {\bibfnamefont {K.}~\bibnamefont {Nishio}}, \bibinfo {author} {\bibfnamefont
  {T.}~\bibnamefont {Ohtsuki}}, \bibinfo {author} {\bibfnamefont
  {H.}~\bibnamefont {Wang}}, \bibinfo {author} {\bibfnamefont {G.}~\bibnamefont
  {Fan}}, \bibinfo {author} {\bibfnamefont {K.}~\bibnamefont {Stopani}},
  \bibinfo {author} {\bibfnamefont {F.}~\bibnamefont {Suzaki}}, \bibinfo
  {author} {\bibfnamefont {K.}~\bibnamefont {Hirose}}, \bibinfo {author}
  {\bibfnamefont {M.}~\bibnamefont {Inagaki}}, \bibinfo {author} {\bibfnamefont
  {Y.-W.}\ \bibnamefont {Lui}}, \bibinfo {author} {\bibfnamefont
  {T.}~\bibnamefont {Ari-izumi}}, \bibinfo {author} {\bibfnamefont
  {S.}~\bibnamefont {Miyamoto}}, \bibinfo {author} {\bibfnamefont
  {T.}~\bibnamefont {Otsuka}},\ and\ \bibinfo {author} {\bibfnamefont
  {H.}~\bibnamefont {Utsunomiya}},\ }\href
  {https://doi.org/10.1103/PhysRevC.109.044602} {\bibfield  {journal} {\bibinfo
   {journal} {Phys. Rev. C}\ }\textbf {\bibinfo {volume} {109}},\ \bibinfo
  {pages} {044602} (\bibinfo {year} {2024})}\BibitemShut {NoStop}%
\bibitem [{\citenamefont {Schmidt}\ \emph {et~al.}(2016)\citenamefont
  {Schmidt}, \citenamefont {Jurado}, \citenamefont {Amouroux},\ and\
  \citenamefont {Schmitt}}]{schGEF}%
  \BibitemOpen
  \bibfield  {author} {\bibinfo {author} {\bibfnamefont {K.-H.}\ \bibnamefont
  {Schmidt}}, \bibinfo {author} {\bibfnamefont {B.}~\bibnamefont {Jurado}},
  \bibinfo {author} {\bibfnamefont {C.}~\bibnamefont {Amouroux}},\ and\
  \bibinfo {author} {\bibfnamefont {C.}~\bibnamefont {Schmitt}},\ }\href
  {https://doi.org/10.1016/j.nds.2015.12.009} {\bibfield  {journal} {\bibinfo
  {journal} {Nucl. Data Sheets}\ }\textbf {\bibinfo {volume} {131}},\ \bibinfo
  {pages} {107} (\bibinfo {year} {2016})},\ \bibinfo {note} {version
  2023}\BibitemShut {NoStop}%
\bibitem [{\citenamefont {Koning}\ \emph {et~al.}(2023)\citenamefont {Koning},
  \citenamefont {Hilaire},\ and\ \citenamefont {Goriely}}]{TALYS}%
  \BibitemOpen
  \bibfield  {author} {\bibinfo {author} {\bibfnamefont {A.}~\bibnamefont
  {Koning}}, \bibinfo {author} {\bibfnamefont {S.}~\bibnamefont {Hilaire}},\
  and\ \bibinfo {author} {\bibfnamefont {S.}~\bibnamefont {Goriely}},\ }\href
  {https://doi.org/10.1140/epja/s10050-023-01034-3} {\bibfield  {journal}
  {\bibinfo  {journal} {The European Physical Journal A}\ }\textbf {\bibinfo
  {volume} {59}},\ \bibinfo {pages} {131} (\bibinfo {year} {2023})}\BibitemShut
  {NoStop}%
\bibitem [{\citenamefont {Tanaka}\ \emph {et~al.}(2022)\citenamefont {Tanaka},
  \citenamefont {Hirose}, \citenamefont {Nishio}, \citenamefont {Kean},
  \citenamefont {Makii}, \citenamefont {Orlandi}, \citenamefont {Tsukada},\
  and\ \citenamefont {Aritomo}}]{Tanaka22}%
  \BibitemOpen
  \bibfield  {author} {\bibinfo {author} {\bibfnamefont {S.}~\bibnamefont
  {Tanaka}}, \bibinfo {author} {\bibfnamefont {K.}~\bibnamefont {Hirose}},
  \bibinfo {author} {\bibfnamefont {K.}~\bibnamefont {Nishio}}, \bibinfo
  {author} {\bibfnamefont {K.~R.}\ \bibnamefont {Kean}}, \bibinfo {author}
  {\bibfnamefont {H.}~\bibnamefont {Makii}}, \bibinfo {author} {\bibfnamefont
  {R.}~\bibnamefont {Orlandi}}, \bibinfo {author} {\bibfnamefont
  {K.}~\bibnamefont {Tsukada}},\ and\ \bibinfo {author} {\bibfnamefont
  {Y.}~\bibnamefont {Aritomo}},\ }\href
  {https://doi.org/10.1103/PhysRevC.105.L021602} {\bibfield  {journal}
  {\bibinfo  {journal} {Phys. Rev. C}\ }\textbf {\bibinfo {volume} {105}},\
  \bibinfo {pages} {L021602} (\bibinfo {year} {2022})}\BibitemShut {NoStop}%
\bibitem [{\citenamefont {Simpson}\ \emph {et~al.}(2000)\citenamefont
  {Simpson}, \citenamefont {Azaiez}, \citenamefont {de~France}, \citenamefont
  {Fouan}, \citenamefont {Gerl}, \citenamefont {Julin}, \citenamefont {Korten},
  \citenamefont {Nolan}, \citenamefont {Nyako}, \citenamefont {Sletten},\ and\
  \citenamefont {Walker}}]{EXOGAM}%
  \BibitemOpen
  \bibfield  {author} {\bibinfo {author} {\bibfnamefont {J.}~\bibnamefont
  {Simpson}}, \bibinfo {author} {\bibfnamefont {F.}~\bibnamefont {Azaiez}},
  \bibinfo {author} {\bibfnamefont {G.}~\bibnamefont {de~France}}, \bibinfo
  {author} {\bibfnamefont {G.}~\bibnamefont {Fouan}}, \bibinfo {author}
  {\bibfnamefont {J.}~\bibnamefont {Gerl}}, \bibinfo {author} {\bibfnamefont
  {R.}~\bibnamefont {Julin}}, \bibinfo {author} {\bibfnamefont
  {W.}~\bibnamefont {Korten}}, \bibinfo {author} {\bibfnamefont
  {P.}~\bibnamefont {Nolan}}, \bibinfo {author} {\bibfnamefont
  {B.}~\bibnamefont {Nyako}}, \bibinfo {author} {\bibfnamefont
  {G.}~\bibnamefont {Sletten}},\ and\ \bibinfo {author} {\bibfnamefont
  {P.}~\bibnamefont {Walker}},\ }\href
  {https://in2p3.hal.science/in2p3-00438782} {\bibfield  {journal} {\bibinfo
  {journal} {{Acta Physica Hungarica New Series-Heavy ion}}\ }\textbf {\bibinfo
  {volume} {11}},\ \bibinfo {pages} {159} (\bibinfo {year} {2000})}\BibitemShut
  {NoStop}%
\bibitem [{\citenamefont {Zielinska}\ \emph {et~al.}(2016)\citenamefont
  {Zielinska}, \citenamefont {Gaffney}, \citenamefont {Wrzosek-Lipska},
  \citenamefont {Clement}, \citenamefont {Grahn}, \citenamefont {Kesteloot},
  \citenamefont {Napiorkowski}, \citenamefont {Pakarinen}, \citenamefont
  {Van~Duppen},\ and\ \citenamefont {Warr}}]{GOSIA}%
  \BibitemOpen
  \bibfield  {author} {\bibinfo {author} {\bibfnamefont {M.}~\bibnamefont
  {Zielinska}}, \bibinfo {author} {\bibfnamefont {L.~P.}\ \bibnamefont
  {Gaffney}}, \bibinfo {author} {\bibfnamefont {K.}~\bibnamefont
  {Wrzosek-Lipska}}, \bibinfo {author} {\bibfnamefont {E.}~\bibnamefont
  {Clement}}, \bibinfo {author} {\bibfnamefont {T.}~\bibnamefont {Grahn}},
  \bibinfo {author} {\bibfnamefont {N.}~\bibnamefont {Kesteloot}}, \bibinfo
  {author} {\bibfnamefont {P.}~\bibnamefont {Napiorkowski}}, \bibinfo {author}
  {\bibfnamefont {J.}~\bibnamefont {Pakarinen}}, \bibinfo {author}
  {\bibfnamefont {P.}~\bibnamefont {Van~Duppen}},\ and\ \bibinfo {author}
  {\bibfnamefont {N.}~\bibnamefont {Warr}},\ }\href
  {https://doi.org/10.1140/epja/i2016-16099-8} {\bibfield  {journal} {\bibinfo
  {journal} {The European Physical Journal A}\ }\textbf {\bibinfo {volume}
  {52}},\ \bibinfo {pages} {99} (\bibinfo {year} {2016})}\BibitemShut {NoStop}%
\bibitem [{\citenamefont {Litaize}\ \emph {et~al.}(2015)\citenamefont
  {Litaize}, \citenamefont {Serot},\ and\ \citenamefont {Berge}}]{fifrelin}%
  \BibitemOpen
  \bibfield  {author} {\bibinfo {author} {\bibfnamefont {O.}~\bibnamefont
  {Litaize}}, \bibinfo {author} {\bibfnamefont {O.}~\bibnamefont {Serot}},\
  and\ \bibinfo {author} {\bibfnamefont {L.}~\bibnamefont {Berge}},\ }\href
  {https://doi.org/10.1140/epja/i2015-15177-9} {\bibfield  {journal} {\bibinfo
  {journal} {The European Physical Journal A}\ }\textbf {\bibinfo {volume}
  {51}},\ \bibinfo {pages} {177} (\bibinfo {year} {2015})}\BibitemShut
  {NoStop}%
\bibitem [{\citenamefont {Kellett}\ \emph {et~al.}(2009)\citenamefont
  {Kellett}, \citenamefont {Besillon},\ and\ \citenamefont {Mills}}]{JEFF}%
  \BibitemOpen
  \bibfield  {author} {\bibinfo {author} {\bibfnamefont {M.~A.}\ \bibnamefont
  {Kellett}}, \bibinfo {author} {\bibfnamefont {O.}~\bibnamefont {Besillon}},\
  and\ \bibinfo {author} {\bibfnamefont {R.~W.}\ \bibnamefont {Mills}},\ }\href
  {https://www.oecd-nea.org/dbdata/nds_jefreports/jefreport-20/nea6287-jeff-20.pdf}
  {\bibinfo {title} {{The JEFF-3.1/-3.1.1 radioactive decay data and fission
  yields sub-libraries, OECD Nuclear Energy Agency Report No. 6287}}} (\bibinfo
  {year} {2009})\BibitemShut {NoStop}%
\bibitem [{\citenamefont {Wilson}\ \emph {et~al.}(2021)\citenamefont {Wilson},
  \citenamefont {Thisse},\ and\ \citenamefont {Lebois}}]{WilsonNature}%
  \BibitemOpen
  \bibfield  {author} {\bibinfo {author} {\bibfnamefont {J.}~\bibnamefont
  {Wilson}}, \bibinfo {author} {\bibfnamefont {D.}~\bibnamefont {Thisse}},\
  and\ \bibinfo {author} {\bibfnamefont {M.~e.~a.}\ \bibnamefont {Lebois}},\
  }\href {https://doi.org/10.1038/s41586-021-03304-w} {\bibfield  {journal}
  {\bibinfo  {journal} {Nature}\ }\textbf {\bibinfo {volume} {590}},\ \bibinfo
  {pages} {566} (\bibinfo {year} {2021})}\BibitemShut {NoStop}%
\bibitem [{\citenamefont {Shrivastava}\ \emph {et~al.}(2009)\citenamefont
  {Shrivastava}, \citenamefont {Caama\~no}, \citenamefont {Rejmund},
  \citenamefont {Navin}, \citenamefont {Rejmund}, \citenamefont {Schmidt},
  \citenamefont {Lemasson}, \citenamefont {Schmitt}, \citenamefont {Gaudefroy},
  \citenamefont {Sieja}, \citenamefont {Audouin}, \citenamefont {Bacri},
  \citenamefont {Barreau}, \citenamefont {Benlliure}, \citenamefont
  {Casarejos}, \citenamefont {Derkx}, \citenamefont {Fern\'andez-Dom\'inguez},
  \citenamefont {Golabek}, \citenamefont {Jurado}, \citenamefont {Roger},\ and\
  \citenamefont {Taieb}}]{Shrivastava09}%
  \BibitemOpen
  \bibfield  {author} {\bibinfo {author} {\bibfnamefont {A.}~\bibnamefont
  {Shrivastava}}, \bibinfo {author} {\bibfnamefont {M.}~\bibnamefont
  {Caama\~no}}, \bibinfo {author} {\bibfnamefont {M.}~\bibnamefont {Rejmund}},
  \bibinfo {author} {\bibfnamefont {A.}~\bibnamefont {Navin}}, \bibinfo
  {author} {\bibfnamefont {F.}~\bibnamefont {Rejmund}}, \bibinfo {author}
  {\bibfnamefont {K.~H.}\ \bibnamefont {Schmidt}}, \bibinfo {author}
  {\bibfnamefont {A.}~\bibnamefont {Lemasson}}, \bibinfo {author}
  {\bibfnamefont {C.}~\bibnamefont {Schmitt}}, \bibinfo {author} {\bibfnamefont
  {L.}~\bibnamefont {Gaudefroy}}, \bibinfo {author} {\bibfnamefont
  {K.}~\bibnamefont {Sieja}}, \bibinfo {author} {\bibfnamefont
  {L.}~\bibnamefont {Audouin}}, \bibinfo {author} {\bibfnamefont {C.~O.}\
  \bibnamefont {Bacri}}, \bibinfo {author} {\bibfnamefont {G.}~\bibnamefont
  {Barreau}}, \bibinfo {author} {\bibfnamefont {J.}~\bibnamefont {Benlliure}},
  \bibinfo {author} {\bibfnamefont {E.}~\bibnamefont {Casarejos}}, \bibinfo
  {author} {\bibfnamefont {X.}~\bibnamefont {Derkx}}, \bibinfo {author}
  {\bibfnamefont {B.}~\bibnamefont {Fern\'andez-Dom\'inguez}}, \bibinfo
  {author} {\bibfnamefont {C.}~\bibnamefont {Golabek}}, \bibinfo {author}
  {\bibfnamefont {B.}~\bibnamefont {Jurado}}, \bibinfo {author} {\bibfnamefont
  {T.}~\bibnamefont {Roger}},\ and\ \bibinfo {author} {\bibfnamefont
  {J.}~\bibnamefont {Taieb}},\ }\href
  {https://doi.org/10.1103/PhysRevC.80.051305} {\bibfield  {journal} {\bibinfo
  {journal} {Phys. Rev. C}\ }\textbf {\bibinfo {volume} {80}},\ \bibinfo
  {pages} {051305} (\bibinfo {year} {2009})}\BibitemShut {NoStop}%
\bibitem [{\citenamefont {Navin}\ \emph {et~al.}(2014)\citenamefont {Navin},
  \citenamefont {Rejmund}, \citenamefont {Schmitt}, \citenamefont
  {Bhattacharyya}, \citenamefont {Lhersonneau}, \citenamefont {Isacker},
  \citenamefont {no}, \citenamefont {Cl\'ement}, \citenamefont {Delaune},
  \citenamefont {Farget}, \citenamefont {de~France},\ and\ \citenamefont
  {Jacquot}}]{Navin14}%
  \BibitemOpen
  \bibfield  {author} {\bibinfo {author} {\bibfnamefont {A.}~\bibnamefont
  {Navin}}, \bibinfo {author} {\bibfnamefont {M.}~\bibnamefont {Rejmund}},
  \bibinfo {author} {\bibfnamefont {C.}~\bibnamefont {Schmitt}}, \bibinfo
  {author} {\bibfnamefont {S.}~\bibnamefont {Bhattacharyya}}, \bibinfo {author}
  {\bibfnamefont {G.}~\bibnamefont {Lhersonneau}}, \bibinfo {author}
  {\bibfnamefont {P.~V.}\ \bibnamefont {Isacker}}, \bibinfo {author}
  {\bibfnamefont {M.~C.}\ \bibnamefont {no}}, \bibinfo {author} {\bibfnamefont
  {E.}~\bibnamefont {Cl\'ement}}, \bibinfo {author} {\bibfnamefont
  {O.}~\bibnamefont {Delaune}}, \bibinfo {author} {\bibfnamefont
  {F.}~\bibnamefont {Farget}}, \bibinfo {author} {\bibfnamefont
  {G.}~\bibnamefont {de~France}},\ and\ \bibinfo {author} {\bibfnamefont
  {B.}~\bibnamefont {Jacquot}},\ }\href
  {https://doi.org/https://doi.org/10.1016/j.physletb.2013.11.024} {\bibfield
  {journal} {\bibinfo  {journal} {Physics Letters B}\ }\textbf {\bibinfo
  {volume} {728}},\ \bibinfo {pages} {136} (\bibinfo {year}
  {2014})}\BibitemShut {NoStop}%
\bibitem [{\citenamefont {Carjan}\ \emph {et~al.}(2007)\citenamefont {Carjan},
  \citenamefont {Talou},\ and\ \citenamefont {Serot}}]{carjan07}%
  \BibitemOpen
  \bibfield  {author} {\bibinfo {author} {\bibfnamefont {N.}~\bibnamefont
  {Carjan}}, \bibinfo {author} {\bibfnamefont {P.}~\bibnamefont {Talou}},\ and\
  \bibinfo {author} {\bibfnamefont {O.}~\bibnamefont {Serot}},\ }\href
  {https://doi.org/https://doi.org/10.1016/j.nuclphysa.2007.05.006} {\bibfield
  {journal} {\bibinfo  {journal} {Nuclear Physics A}\ }\textbf {\bibinfo
  {volume} {792}},\ \bibinfo {pages} {102} (\bibinfo {year}
  {2007})}\BibitemShut {NoStop}%
\bibitem [{\citenamefont {Capote}\ \emph {et~al.}(2016)\citenamefont {Capote},
  \citenamefont {Carjan},\ and\ \citenamefont {Chiba}}]{capote16}%
  \BibitemOpen
  \bibfield  {author} {\bibinfo {author} {\bibfnamefont {R.}~\bibnamefont
  {Capote}}, \bibinfo {author} {\bibfnamefont {N.}~\bibnamefont {Carjan}},\
  and\ \bibinfo {author} {\bibfnamefont {S.}~\bibnamefont {Chiba}},\ }\href
  {https://doi.org/10.1103/PhysRevC.93.024609} {\bibfield  {journal} {\bibinfo
  {journal} {Phys. Rev. C}\ }\textbf {\bibinfo {volume} {93}},\ \bibinfo
  {pages} {024609} (\bibinfo {year} {2016})}\BibitemShut {NoStop}%
\bibitem [{\citenamefont {Abdurrahman}\ \emph {et~al.}(2024)\citenamefont
  {Abdurrahman}, \citenamefont {Kafker}, \citenamefont {Bulgac},\ and\
  \citenamefont {Stetcu}}]{Abdurrahman24}%
  \BibitemOpen
  \bibfield  {author} {\bibinfo {author} {\bibfnamefont {I.}~\bibnamefont
  {Abdurrahman}}, \bibinfo {author} {\bibfnamefont {M.}~\bibnamefont {Kafker}},
  \bibinfo {author} {\bibfnamefont {A.}~\bibnamefont {Bulgac}},\ and\ \bibinfo
  {author} {\bibfnamefont {I.}~\bibnamefont {Stetcu}},\ }\href
  {https://doi.org/10.1103/PhysRevLett.132.242501} {\bibfield  {journal}
  {\bibinfo  {journal} {Phys. Rev. Lett.}\ }\textbf {\bibinfo {volume} {132}},\
  \bibinfo {pages} {242501} (\bibinfo {year} {2024})}\BibitemShut {NoStop}%
\bibitem [{\citenamefont {Cannarozzo}\ \emph {et~al.}(2025)\citenamefont
  {Cannarozzo}, \citenamefont {Pomp}, \citenamefont {Solders}, \citenamefont
  {Al-Adili}, \citenamefont {Gao}, \citenamefont {Lantz}, \citenamefont
  {Penttil\"a}, \citenamefont {Kankainen}, \citenamefont {Moore}, \citenamefont
  {Eronen}, \citenamefont {Ge}, \citenamefont {Ruotsalainen}, \citenamefont
  {Mougeot}, \citenamefont {Virtanen}, \citenamefont {Jaries}, \citenamefont
  {Stryjczyk},\ and\ \citenamefont {Raggio}}]{cannarozzo24}%
  \BibitemOpen
  \bibfield  {author} {\bibinfo {author} {\bibfnamefont {S.}~\bibnamefont
  {Cannarozzo}}, \bibinfo {author} {\bibfnamefont {S.}~\bibnamefont {Pomp}},
  \bibinfo {author} {\bibfnamefont {A.}~\bibnamefont {Solders}}, \bibinfo
  {author} {\bibfnamefont {A.}~\bibnamefont {Al-Adili}}, \bibinfo {author}
  {\bibfnamefont {Z.}~\bibnamefont {Gao}}, \bibinfo {author} {\bibfnamefont
  {M.}~\bibnamefont {Lantz}}, \bibinfo {author} {\bibfnamefont
  {H.}~\bibnamefont {Penttil\"a}}, \bibinfo {author} {\bibfnamefont
  {A.}~\bibnamefont {Kankainen}}, \bibinfo {author} {\bibfnamefont
  {I.}~\bibnamefont {Moore}}, \bibinfo {author} {\bibfnamefont
  {T.}~\bibnamefont {Eronen}}, \bibinfo {author} {\bibfnamefont
  {Z.}~\bibnamefont {Ge}}, \bibinfo {author} {\bibfnamefont {J.}~\bibnamefont
  {Ruotsalainen}}, \bibinfo {author} {\bibfnamefont {M.}~\bibnamefont
  {Mougeot}}, \bibinfo {author} {\bibfnamefont {V.}~\bibnamefont {Virtanen}},
  \bibinfo {author} {\bibfnamefont {A.}~\bibnamefont {Jaries}}, \bibinfo
  {author} {\bibfnamefont {M.}~\bibnamefont {Stryjczyk}},\ and\ \bibinfo
  {author} {\bibfnamefont {A.}~\bibnamefont {Raggio}},\ }\href
  {https://doi.org/10.1103/PhysRevC.111.L031601} {\bibfield  {journal}
  {\bibinfo  {journal} {Phys. Rev. C}\ }\textbf {\bibinfo {volume} {111}},\
  \bibinfo {pages} {L031601} (\bibinfo {year} {2025})}\BibitemShut {NoStop}%
\bibitem [{\citenamefont {Duderstadt}\ and\ \citenamefont
  {Hamilton}(1976)}]{Duderstadt}%
  \BibitemOpen
  \bibfield  {author} {\bibinfo {author} {\bibfnamefont {J.}~\bibnamefont
  {Duderstadt}}\ and\ \bibinfo {author} {\bibfnamefont {L.}~\bibnamefont
  {Hamilton}},\ }\href@noop {} {\emph {\bibinfo {title} {``Nuclear Reactor
  Analysis''}}},\ edited by\ \bibinfo {editor} {\bibnamefont {{John Wiley \&
  Sons}}}\ (\bibinfo {year} {1976})\BibitemShut {NoStop}%
\bibitem [{\citenamefont {Marguet}(2017)}]{Marguet2017}%
  \BibitemOpen
  \bibfield  {author} {\bibinfo {author} {\bibfnamefont {S.}~\bibnamefont
  {Marguet}},\ }\href@noop {} {\emph {\bibinfo {title} {``The Physics of
  Nuclear Reactors''}}},\ edited by\ \bibinfo {editor} {\bibnamefont
  {Springer}}\ (\bibinfo {year} {2017})\BibitemShut {NoStop}%
\bibitem [{\citenamefont {Doligez}\ \emph {et~al.}(2017)\citenamefont
  {Doligez}, \citenamefont {Bouneau}, \citenamefont {David}, \citenamefont
  {Ernoult}, \citenamefont {Zakari-Issoufou}, \citenamefont {Thiolli\`ere},
  \citenamefont {Bidaud}, \citenamefont {M\'eplan}, \citenamefont {Nuttin},\
  and\ \citenamefont {Capellan}}]{Doligez}%
  \BibitemOpen
  \bibfield  {author} {\bibinfo {author} {\bibfnamefont {X.}~\bibnamefont
  {Doligez}}, \bibinfo {author} {\bibfnamefont {S.}~\bibnamefont {Bouneau}},
  \bibinfo {author} {\bibfnamefont {S.}~\bibnamefont {David}}, \bibinfo
  {author} {\bibfnamefont {M.}~\bibnamefont {Ernoult}}, \bibinfo {author}
  {\bibfnamefont {A.-A.}\ \bibnamefont {Zakari-Issoufou}}, \bibinfo {author}
  {\bibfnamefont {N.}~\bibnamefont {Thiolli\`ere}}, \bibinfo {author}
  {\bibfnamefont {A.}~\bibnamefont {Bidaud}}, \bibinfo {author} {\bibfnamefont
  {O.}~\bibnamefont {M\'eplan}}, \bibinfo {author} {\bibfnamefont
  {A.}~\bibnamefont {Nuttin}},\ and\ \bibinfo {author} {\bibfnamefont
  {N.}~\bibnamefont {Capellan}},\ }\href
  {https://doi.org/10.1016/j.crhy.2017.10.004} {\bibfield  {journal} {\bibinfo
  {journal} {Comptes Rendus. Physique}\ }\textbf {\bibinfo {volume} {18}},\
  \bibinfo {pages} {372} (\bibinfo {year} {2017})}\BibitemShut {NoStop}%
\end{thebibliography}%
\bibliographystyle{apsrev4-2}

\end{document}